This is an author's version of a paper published in the Journal of Network and Computer Applications: Ana Larrañaga, M. Carmen Lucas-Estañ, Sandra Lagén, Zoraze Ali, Imanol Martinez, Javier Gozalvez, "An open-source implementation and validation of 5G NR configured grant for URLLC in ns-3 5G LENA: A scheduling case study in industry 4.0 scenarios", *Journal of Network and Computer Applications*, Volume 215, 2023, 103638, https://doi.org/10.1016/j.jnca.2023.103638. You can access the published version at https://www.sciencedirect.com/science/article/abs/pii/S1084804523000577.

# An open-source implementation and validation of 5G NR Configured Grant for URLLC in ns-3 5G LENA: a scheduling case study in Industry 4.0 scenarios

Ana Larrañaga[a], M. Carmen Lucas-Estañ[b],

Sandra Lagen[c], Zoraze Ali[c], Imanol Martinez[a], Javier Gozalvez[b]

[a]*HW and Communication Systems Area, Ikerlan Technology Research Centre, Mondragón 20500, Spain*
[b]*UWICORE Laboratory, Universidad Miguel Hernández de Elche (UMH), Elche 03202, Spain*
[c]*Centre Tecnològic de Telecomunicacions de Catalunya (CTTC/CERCA), Castelldefels, Barcelona, Spain*

**Abstract** Factories are undergoing a digital transformation towards more cost-efficient, zero-defect manufacturing. The digitalized factories require communication networks capable of satisfying their strict latency and reliability demands. 5G and beyond networks are being designed to efficiently support services demanding Ultra-Reliable and Low Latency Communications (URLLC). At the MAC level, the use of dynamic scheduling for uplink transmissions entails a non-negligible latency introduced by the signaling messages exchanged to request and inform about the radio resources allocated for each packet transmission. To reduce the transmission latency, 5G defines Configured Grant (CG) for UL transmissions that pre-allocates radio resources to the User Equipments (UEs) and eliminates the need for requesting resources for each transmission. In this context, the availability of 5G NR simulation tools that accurately implement all 5G NR functionalities, and in particular, the technological enablers introduced in 5G NR to support URLLC, is key to analyze the capability of 5G and beyond networks to support time-critical services and research on new solutions. The availability and access to such tools are limited, and to the best of the authors' knowledge, there are currently no open-source 5G NR simulators that implement configured grant in 5G NR. To overcome this issue, this work presents the first implementation of configured grant in an open-source 5G NR simulator. In particular, configured grant has been implemented in the ns-3 5G-LENA system-level simulator, and it is publicly available. To accurately model the flexibility of 5G NR, we have also improved the implementation of Orthogonal Frequency Division Multiple Access (OFDMA) access mode in 5G-LENA according to 5G NR. To validate the implementation of CG and analyze the capability of 5G NR to support time-critical services, we analyze the latency performance that can be achieved using CG with different scheduling policies in Industry 4.0 scenarios. The results show that the latency values achieved with CG in 5G-LENA match with those reported by previous analytical studies. In addition, this study shows the importance of efficiently using radio resources to reduce the latency experienced and meet the requirements of critical services.

*Keywords:* 5G, Configured Grant, Grant-free scheduler, scheduling, software simulation, URLLC, ns-3, 5G-LENA

## 1. Introduction

Factories are undergoing a digital transformation towards more cost-efficient, zero-defect manufacturing environments capable of flexibly adapting to changes in production and demand [1]. Industrial applications such as digital twins, motion control, or control-to-control communications demand strict latency and cycle times requirements ranging from milliseconds to microseconds [2] [3]. They require communication networks capable of satisfying their communication requirements. 5G networks are designed to support Ultra-Reliable and Low-Latency Communications (URLLC). In fact, 5G and its future evolution are considered critical enablers for Industry 4.0 paradigm. 5G introduces several mechanisms at the PHY and MAC layers to reduce latency. For example, 5G includes flexible numerologies with different slot durations (from 1 ms to 0.0625 ms) and mini-slot transmissions. In addition to dynamic scheduling, 5G introduces semi-static scheduling to reduce communication latency. With dynamic scheduling, the base station or gNB informs the UEs about the radio resources used for each packet transmitted in Downlink (DL) before the transmission of each packet. When a UE has a packet to transmit in Uplink (UL), the UE requests resources from the gNB, and the gNB replies with a grant and the information about the allocated resources. This process introduces a non-negligible transmission delay, which is higher in UL due to the higher number of messages exchanged between the UE and the gNB. Semi-static scheduling (referred to as Configured Grant or CG for UL transmissions and Semi-Persistent Scheduling or SPS for DL transmissions) pre-allocates radio resources periodically to UEs [4], and UEs can transmit in the pre-allocated resources as soon as they have data to transmit. Semi-static scheduling eliminates the latency introduced by requesting radio resources for each transmission.

Availability and access to accurate simulation tools are critical to analyze the capability of 5G NR to support critical industrial services and research new solutions. However, to the best of the authors' knowledge, there are currently no open-source 5G NR simulators that implement configured grant in 5G NR, which is critical for URLLC services. To overcome this issue, this work presents the first implementation of configured grant in an open-source 5G NR simulator. In particular, configured grant has been implemented in 5G-LENA [5]. 5G-LENA is an open-source discrete-event network simulator of the 5G NR based on ns-3 [6]. 5G-LENA implements most of the main 5G NR functionalities, such as the NR frame structures, the numerologies, or Bandwidth Parts (BWPs). However, 5G-LENA does not implement configured grant scheduling. The code of the configured grant implementation for 5G-LENA is publicly available in [7], and it uses the ns-3 version available in [8].

The latency performance achievable with CG strongly depends on the multiple access scheme since it establishes how radio resources can be shared by the UEs and determines the flexibility of the scheduler to allocate radio resources to the UEs. It is then necessary to accurately model the OFDMA multiple access scheme used in 5G NR in order to analyze the actual capability of 5G NR using CG. 5G-LENA implements several multiple access schemes, but none of them models the flexibility of OFDMA that allows simultaneous radio resource allocations in the time and frequency domain. In this context, we have also implemented OFDMA in 5G-LENA according to 5G NR.

To validate the implementation of CG and analyze the capability of 5G NR to support critical industrial services, we consider a scheduling case study in an Industry 4.0 scenario. In particular, this work analyzes the latency performance that can be achieved using CG with different scheduling policies. The results show that the latency values achieved with CG in 5G-LENA match with those reported by previous analytical studies. In addition, the results highlight the importance of making efficient use of radio resources to reduce the latency experienced and meet the requirements of critical services.

The rest of the paper is organized as follows. 5G New Radio is presented in Section II, and Section III describes the 5G-LENA network simulator. Section IV presents the implementation of CG in 5G-LENA. Section IV also presents the modifications done in 5G-LENA to accurately model OFDMA and the scheduling policies used to evaluate the performance of CG. Section V describes the evaluation scenario and the schemes used as a reference. In Section VI, we derive analytical expressions of the maximum latency experienced with CG and the different scheduling policies. We compare in Section VI the latency achieved analytically and by simulation to validate the implementation of CG. Section VII presents the performance results. Section VIII concludes the paper.

## 2. 5G New Radio

5G uses OFDM (Orthogonal Frequency Division Multiplexing). Radio resources are organized in Resource Blocks (RBs) in the frequency domain and in slots in the time domain (a slot consists of 14 or 12 OFDM symbols when the normal or extended cyclic prefix is used, respectively). A radio resource is composed of an RB, which contains 12 consecutive subcarriers in the frequency domain and a single OFDM symbol in the time domain. While the Subcarrier Spacing (SCS) is fixed at 15 kHz in Long Term Evolution (LTE), 5G defines multiple numerologies $\mu$ that allow the use of different SCSs. A numerology $\mu$ is given by the use of an SCS and slot duration. Numerologies $\mu$ from 0 to 4 use an SCS equal to 15, 30, 60, 120, and 240 kHz SCS, respectively [9]. The slot duration is given by $1/2^{\mu}$ ms, which results in slot durations from 1 ms for $\mu$=0 to 0.0625 ms for $\mu$=4 (in LTE, the 14 symbols-slot duration is always equal to 1 ms). Slots are organized in frames of 10 ms. A slot consists of 14 OFDM symbols in the time domain when the normal Cyclic Prefix (CP) is used (the normal CP can be used with all the numerologies). An extended CP can also be used with $\mu$=2. In this case, a slot consists of 12 OFDM symbols. Numerologies 0, 1, and 2 can be used in the lower frequency range (410 MHz-7.125 GHz), and numerologies 2, 3, and 4 can be used in the higher frequency range (24.25 GHz-52.6 GHz). 5G NR allows transmissions to start at any OFDM symbol within a slot and to use only the number of symbols needed for the transmission. This results in mini-slot transmissions when transmissions only use part of the symbols of a slot (mini-slot transmissions can use between 1 and 13 OFDM symbols in UL and between 2 and 13 OFDM symbols in DL [10]), or full-slot transmissions when all the symbols are used (LTE only considers full-slot transmissions). The introduction of multiple numerologies with higher SCS and lower slot durations, and the use of mini-slot transmissions allow to considerably reduce the latency compared to LTE.

5G can use Frequency-Division Duplex (FDD) or Time-Division Duplex (TDD) modes. On the one hand, TDD provides high flexibility as each slot in a frame can be configured for UL or DL transmissions. Furthermore, a single slot can be split into segments of consecutive symbols that can be used for UL or DL [11]. On the other hand, FDD organizes UL and DL transmissions on separate frequencies. FDD can reduce communication latency since resources are always available simultaneously for UL and DL transmissions.

5G NR uses Low Density Parity Check coding with Quadrature Phase Shift Keying (QPSK), 16 Quadrature Amplitude Modulation (QAM), 64 QAM, or 256 QAM for data channels. 5G NR defines three Modulation and Coding Scheme (MCS) tables that provide a different trade-off between spectrum efficiency and protection against error. MCS tables 1 and 2 in [12] guarantee a Block Error Rate (BLER) of 10%, and MCS table 3 in [12] guarantees a BLER of $10^{-5}$ when used according to the Channel Quality Indicator (CQI).

5G NR uses OFDMA multiple access scheme that allows UEs to share radio resources in the frequency and time domains. With OFDMA, an RB can be used by different UEs in different OFDM symbols, and different RBs in the same OFDM symbol can be assigned to different UEs. As a result, UEs can receive any number of OFDM symbols and RBs. Transmissions in 5G NR can be performed using dynamic or semi-static scheduling in both UL and DL [13]. With dynamic scheduling, the gNB allocates resources dynamically for each packet transmission. When a packet is generated in DL for a UE, the gNB sends a control message to inform the UE about the allocated radio resources. When a UE needs to transmit a packet in UL, the UE sends a scheduling request (SR) to the gNB. The gNB then replies with a grant message to the UE with the information about the radio resources to use for the packet transmission. The UE transmits the packet together with the buffer status report (BSR) to inform the gNB if it has more data to transmit. If this is the case, the gNB will allocate radio resources for a new transmission and will inform the UE. This process is repeated while the UE has pending data to transmit. Dynamic scheduling makes efficient use of radio resources. However, the signaling exchange between the UE and the gNB before the packet transmission can increase the latency of the transmission. This is more critical for UL transmissions since the amount of signaling between the UE and the gNB is higher. This cannot be appropriate for latency-critical industrial applications. Semi-static scheduling can also be used for packet transmission in 5G NR: SPS in DL and CG in UL. CG and SPS pre-assign radio resources periodically to the UEs before the data packets are generated. When a data packet is generated, the packet can be transmitted in the pre-allocated resources. CG and SPS avoid the signaling exchange between the UE and the gNB to request/inform about the allocated radio resources, reducing the transmission latency. The gNB estimates the radio resources necessary to support the UE based on the characteristics of the traffic it will transmit (e.g., data packet size and periodicity). If the gNB has sufficient resources available to support the UE, it allocates the radio resources and informs the UEs about the pre-allocated radio resources and their periodicity using Radio Resource Control (RRC) signaling during the connection setup. A connection is not established if the gNB does not have enough radio resources to satisfy the request of a UE. Two types of CG are defined in [13]: type 1, which activates the configured uplink grant from the moment it is configured via RRC signaling, and type 2, where the grant is activated or deactivated using DL control messages. CG type 1 is illustrated in Figure 1. We should note that if the characteristics of the traffic generated by the UE change (e.g., its size or periodicity), the configured UL grant should be modified, and this can be done through RRC signaling procedures defined in 3rd Generation Partnership Project (3GPP) [14].

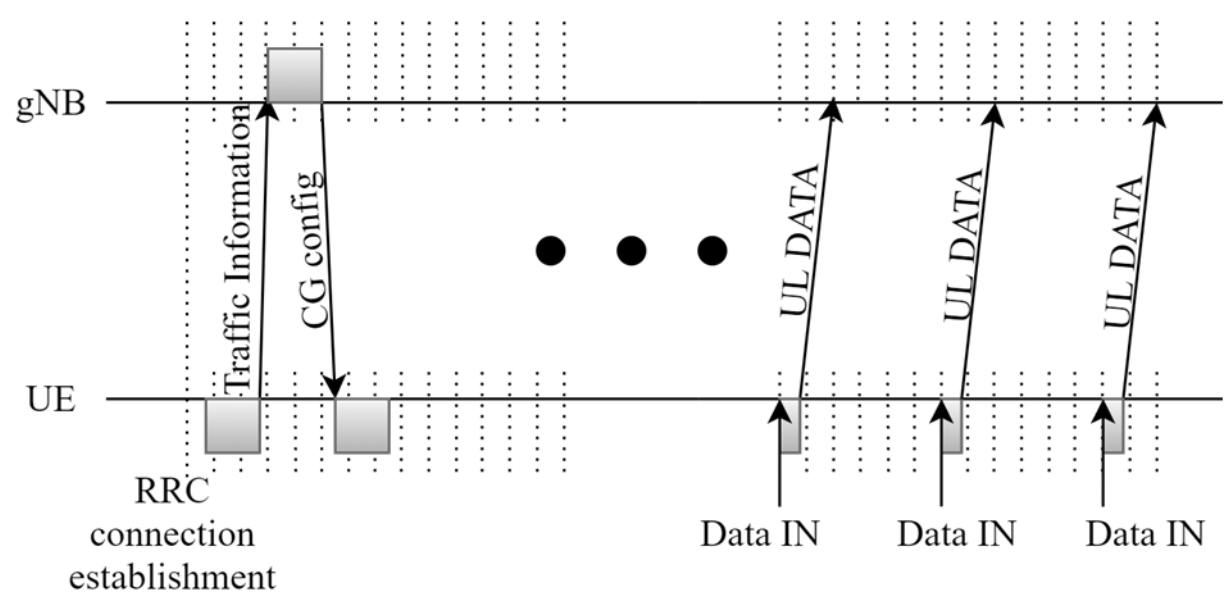

**Figure 1.** Configured grant type 1.

## 3. 5G-LENA network simulator

### 3.1. 5G-LENA overview

5G-LENA is an open-source discrete-event network simulator of the 5G NR [5] implemented over ns-3. ns-3 is an open-source C++ simulation environment for networking research that offers a solid simulation core that supports research on both IP and non-IP-based networks. 5G-LENA is the evolution of LENA that was initially developed to implement the Radio Access Network (RAN) and the core network of LTE. The efforts on the development of 5G-LENA focused first on the RAN. 5G-LENA implements the fundamental PHY-MAC NR features in line with the NR specifications [13], and it has been calibrated in indoor hotspots [5] and outdoor 3GPP reference scenarios [15].

5G-LENA implements two main C++ classes to simulate the functionalities of the gNB (NrGnb class) and the UE (NrUe class) (see Figure 2). For the gNB and the UE, 5G-LENA models the different layers of the protocol stack: PHY, MAC, Radio Link Control (RLC), Packet Data Convergence

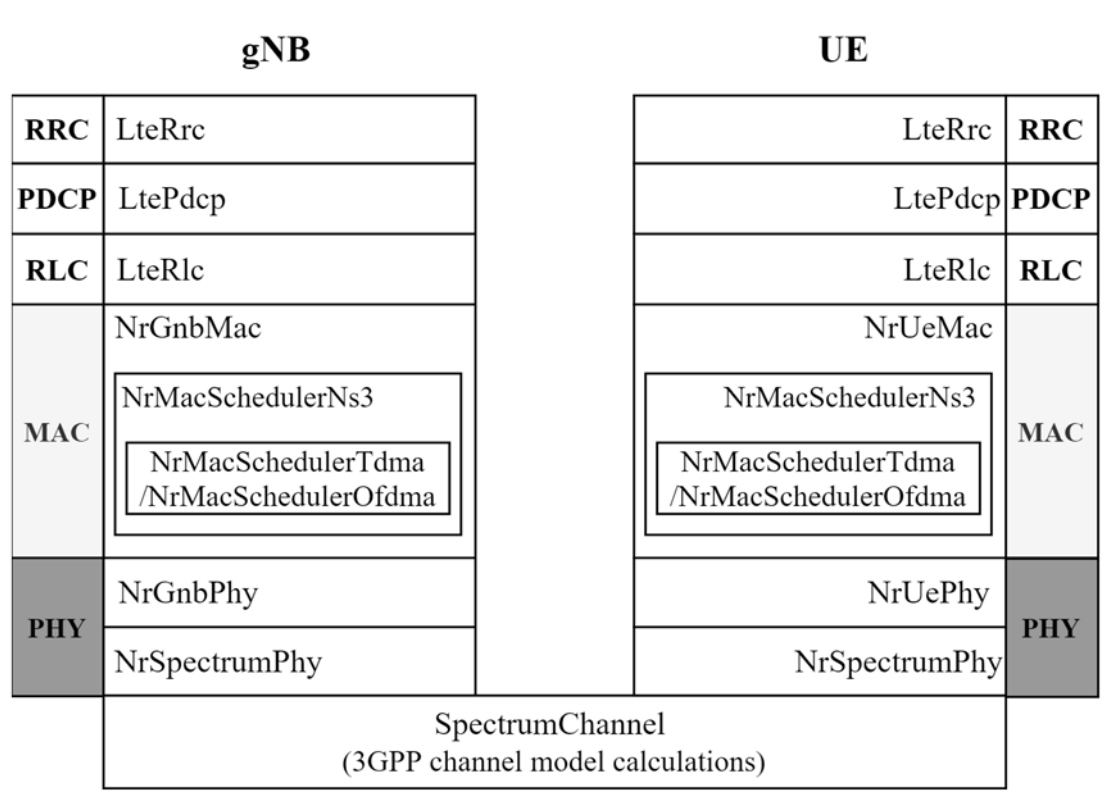

**Figure 2.** Logical representation of the 5G-LENA RAN class.

Protocol (PDCP), and RRC. The upper layers RLC, PDCP, and RRC currently rely on LTE LENA implementation. The PHY and MAC layers implement the main NR features. For example, the PHY layer implements the flexible frame structure defined in [13] for 5G NR. Both TDD and FDD duplexing modes can be configured. In TDD, slots can be flexibly configured for DL or UL transmissions as established by 5G 3GPP standards. 5G-LENA also allows flexible configuration of the number of OFDM symbols within a slot to be used for UL and DL transmissions. The 5G-LENA PHY layer considers the use of the new numerologies defined in [13] for 5G NR, both normal and extended CP, and the different modulation and coding schemes defined in [12]. 5G-LENA provides a realistic implementation of K1 and K2 scheduling timings defined for 3GPP 5G NR. The K1 scheduling timing corresponds to the time between DL data reception at the UE and the corresponding HARQ-ACK feedback transmission to the gNB, while K2 is the delay between UL grant reception at the UE and the corresponding UL data transmission to the gNB. 5G-LENA allows the configuration of such parameters to simulate UEs of different capabilities and processing times [16]. Two beamforming methods have been added, long-term covariance matrix and beam-search, using either ideal beamforming or realistic beamforming based on uplink Sounding Reference Signals [17]. Recently, dual-polarized Multiple-Input Multiple-Output (MIMO) for the downlink has been implemented [18]. 5G-LENA also supports different channel and propagation models according to the 3GPP spatial channel model defined in TR 38.901 [19].

### 3.2. Multiple access and scheduling in 5G-LENA

5G-LENA has also evolved the MAC to support 5G NR. 5G LENA implements three multiple access schemes referred to as 5GL-TDMA, 5GL-OFDMA, and 5GL-OFDMA with variable TTI [20][1]. Figure 3 shows how radio resources are allocated to several UEs when different multiple access schemes are used. It depicts the time-frequency resource grid for a slot of 14 OFDM symbols in time and $R_{BW}$ number of RBs in the frequency domain. Squares in Figure 3 represent an OFDM symbol in the time domain and one RB in the frequency domain. The first and last symbols are reserved for DL and UL control channels. In this example, each UE needs 4 RBs to transmit its packet. Colored squares represent the RBs assigned to UEs for packet transmission. With 5GL-TDMA, UEs access radio resources at different OFDM symbols, and all RBs in a symbol must be assigned to the same UE (see Figure 3.a). Although UEs only require 4 radio resources to transmit their packets, they receive all the RBs ($R_{BW}$) in an OFDM symbol when 5GL-TDMA is used. 5GL-OFDMA is a constrained version of OFDMA defined in 5G NR. 5GL-OFDMA allocates an RB in all the OFDM symbols within a slot to the same UE. This is shown in Figure 3.b. 5G-LENA implements a modified version of 5GL-OFDMA referred to as 5GL-OFDMA with variable TTI. It allows dividing the OFDM symbols within a slot into two or more segments with a different or equal number of OFDM symbols. The radio resources within each of these segments can be accessed by the UEs assigned to the same antenna beam. In each of these segments, 5GL-OFDMA is applied (see Figure 3.c).

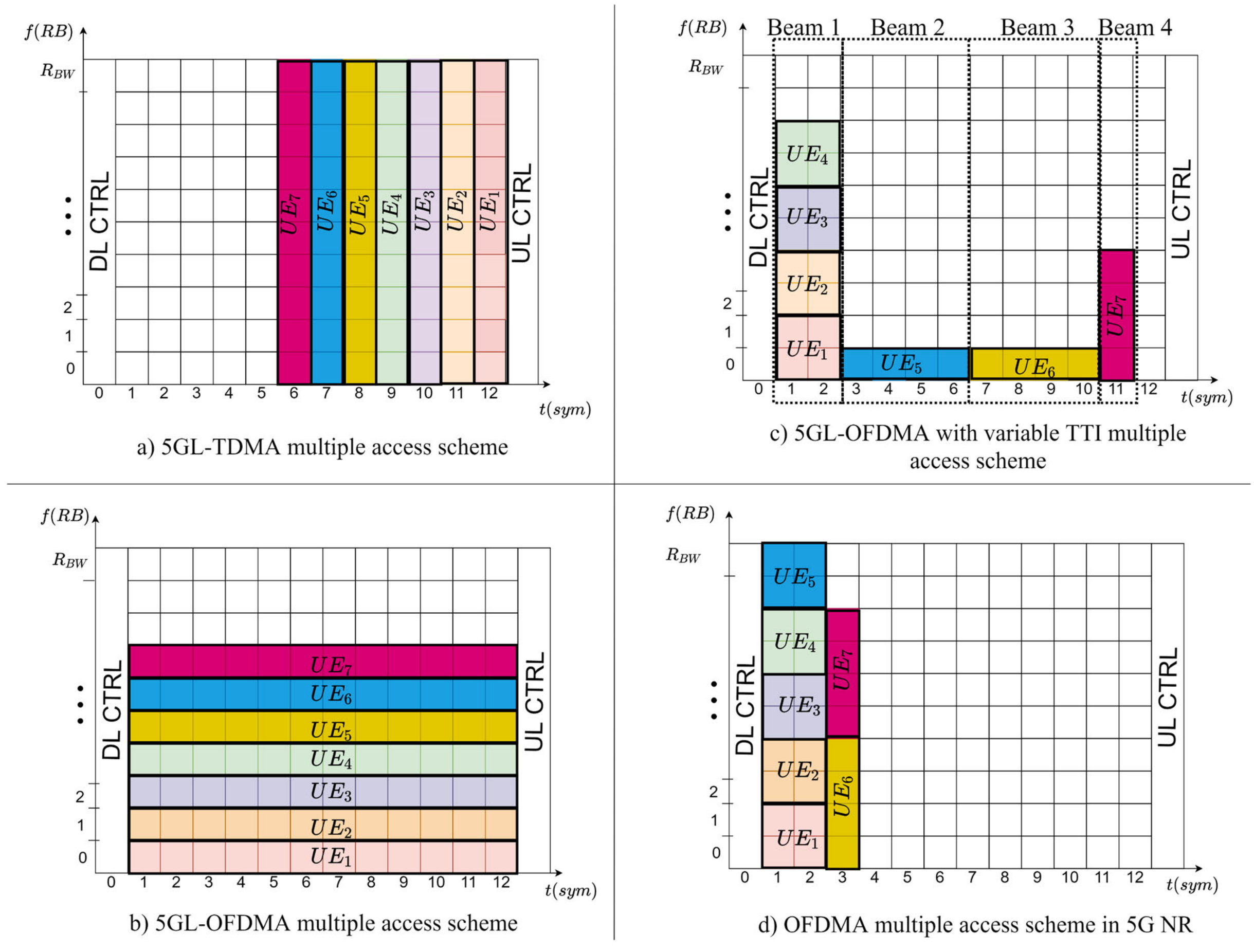

**Figure 3.** Example of radio resource allocation with the different multiple access modes.

[1] Within 5G-LENA, the implemented multiple access schemes are referred to as TDMA, OFDMA and OFDMA with variable TTI. We added the '5GL-' prefix to differentiate between the particular implementation of the multiple access schemes in 5G-LENA and the TDMA and OFDMA multiple access schemes defined in 5G NR (as they may have some differences).

None of these multiple access schemes accurately models the flexibility offered by OFDMA in 5G NR. OFDMA allows allocating different numbers of RBs and OFDM symbols to UEs. An example is shown in Figure 3.d. The 5GL-TDMA and 5GL-OFDMA schemes are constrained to allocate a number of radio resources that is multiple of $R_{BW}$ (the number of available RBs in a particular bandwidth) or $S_{slot}$ (that represents the number of OFDM symbols reserved for UL data transmission within a slot), respectively. In the case of 5GL-OFDMA with variable TTI, the number of radio resources allocated to UEs is multiple of the number of OFDM symbols within a segment. OFDMA in 5G NR allows adjusting the number of assigned resources more accurately to the UEs' demand. This enables a more efficient use of radio resources.

5G-LENA implements dynamic scheduling in DL and UL; semi-static scheduling is not yet implemented in 5G-LENA. Different scheduling policies are currently implemented, such as round-robin, proportional fair, etc., but other policies can also be included. It is important to note that a UE receives a number $r$ of RBs in $s$ consecutive OFDM symbols, i.e., a UE receives $r$ x $s$ radio resources. Due to interference model limitations in 5G-LENA, if two UEs receive RBs in the same OFDM symbol, the two UEs must receive RBs in the same number $s$ of OFDM symbols [5]. Finally, 5G-LENA allocates radio resources for UL transmissions from the last symbol to the first within a slot.

## 4. Implementation of Configured Grant in 5G-LENA

### 4.1. Configured Grant

As presented in previous sections, 5G-LENA does not implement semi-persistent scheduling, which limits its applicability for the study of critical use cases. In this context, 5G-LENA has been extended in this work to simulate UL transmissions using configured grant. We have implemented configured grant Type 1 that configures and stores the UL grant at the session establishment. Configured grant Type 1 avoids potential delays that might be introduced in the activation/deactivation of the configured UL grant. Therefore, it is more suitable for services with stringent latency requirements.

Configured grant has been implemented in 5G-LENA using a state machine that is presented in Figure 4. The state machine is implemented at the UE to manage UL transmissions. INACTIVE_CG is the initial state and is activated for each UE when the simulation starts. The UE informs the gNB about characteristics of the data traffic it intends to transmit, in particular, the size and periodicity of the data packets. To this end, the UE prepares a message with the information about the characteristics of the data traffic and the UE state changes to TO_SEND_TrafficInfo. The UE transmits the prepared message to the gNB in the next slot, and changes to TO_RECEIVE_CG state. When the gNB receives information, it checks the necessary radio resources to support the UE transmissions, and if they are available, it allocates them following the selected scheduling policy. This process is performed within the NrMacSchedulerNs3 class in the DoScheduleUlData function (Figure 5). This function executes the radio resource allocation process for configured UL grant transmissions. Figure 5 illustrates the operation of the DoScheduleUlData function. This function first checks if there are UEs waiting to receive a configured UL grant. If this is the case, the gNB decides the number of OFDM symbols and RBs that should be allocated to each UE. This decision is implemented in AssignULRBG function. Once the decision for radio resource allocation has been made, CreateUlCGConfig function is called to generate the *ConfiguredGrantConfig* message [14] with the configured UL grant and information about the pre-allocated radio resources. This message is a new structure that stores the traffic periodicity, the packet size, the MCS, and the resources that have been allocated to the corresponding UE. The gNB transmits the *ConfiguredGrantConfig* message to the UE. When the UE receives the message, it stores the configured UL grant and changes to ACTIVE_CG state. The UE checks if it is in ACTIVE_CG state (that means it has a configured UL grant) each time a new data packet is generated. If this is the case, the UE changes to SCH_CG_DATA. The UE prepares the data packet to be transmitted and changes to ACTIVE_CG state. The packet is then transmitted to the gNB on the pre-allocated resources. The gNB waits to receive data packets from the UE on the pre-allocated resources. It is important to note that the implementation of configured grant in 5G-LENA has also entailed the extension of several functions at the PHY layer class to allow periodic scheduling (for example, StartSlot function of the UE PHY class has been extended to store the *ConfiguredGrantConfig* message in the UE PHY layer in a periodic way).

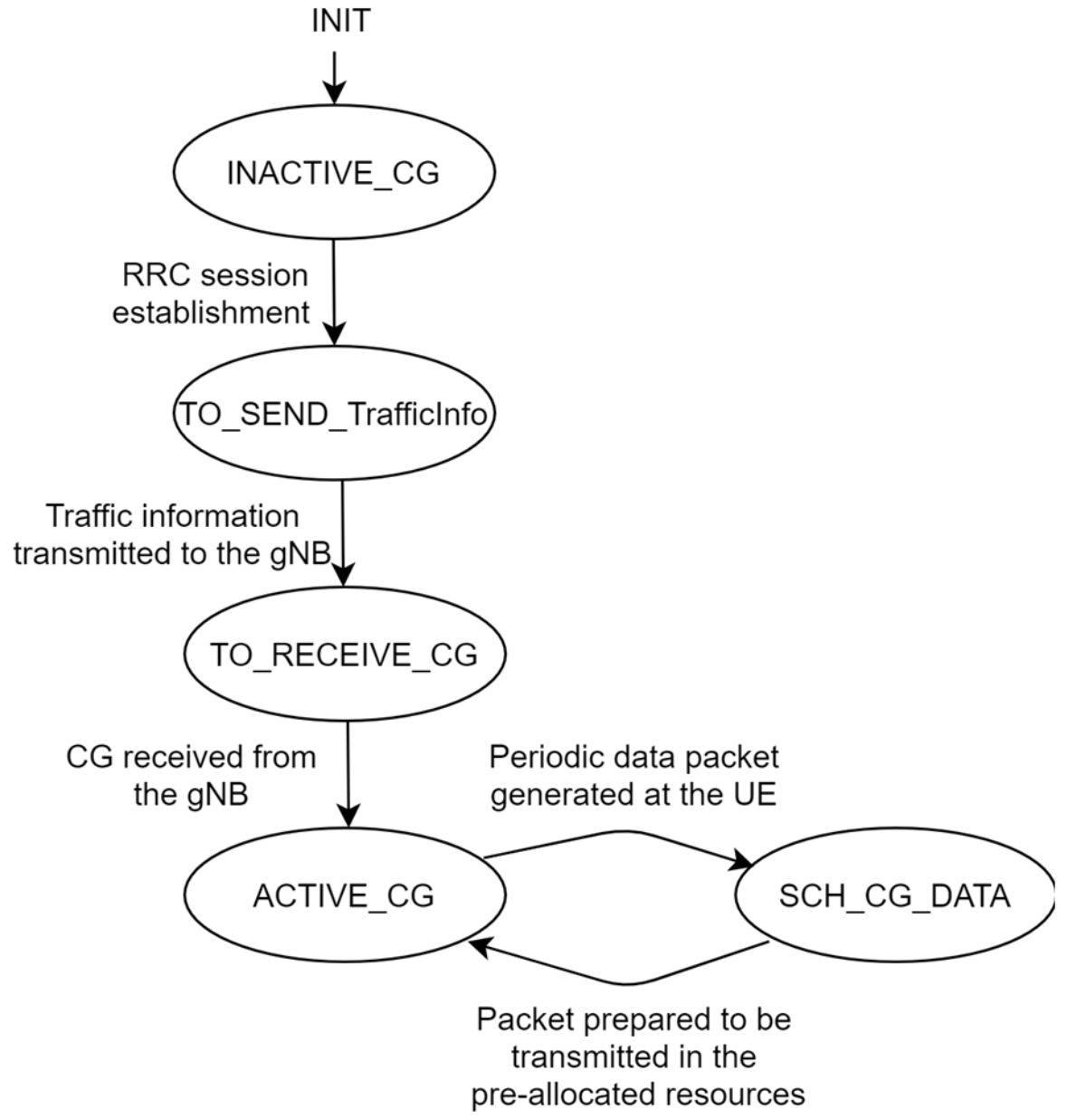

**Figure 4.** Configured grant state machine.

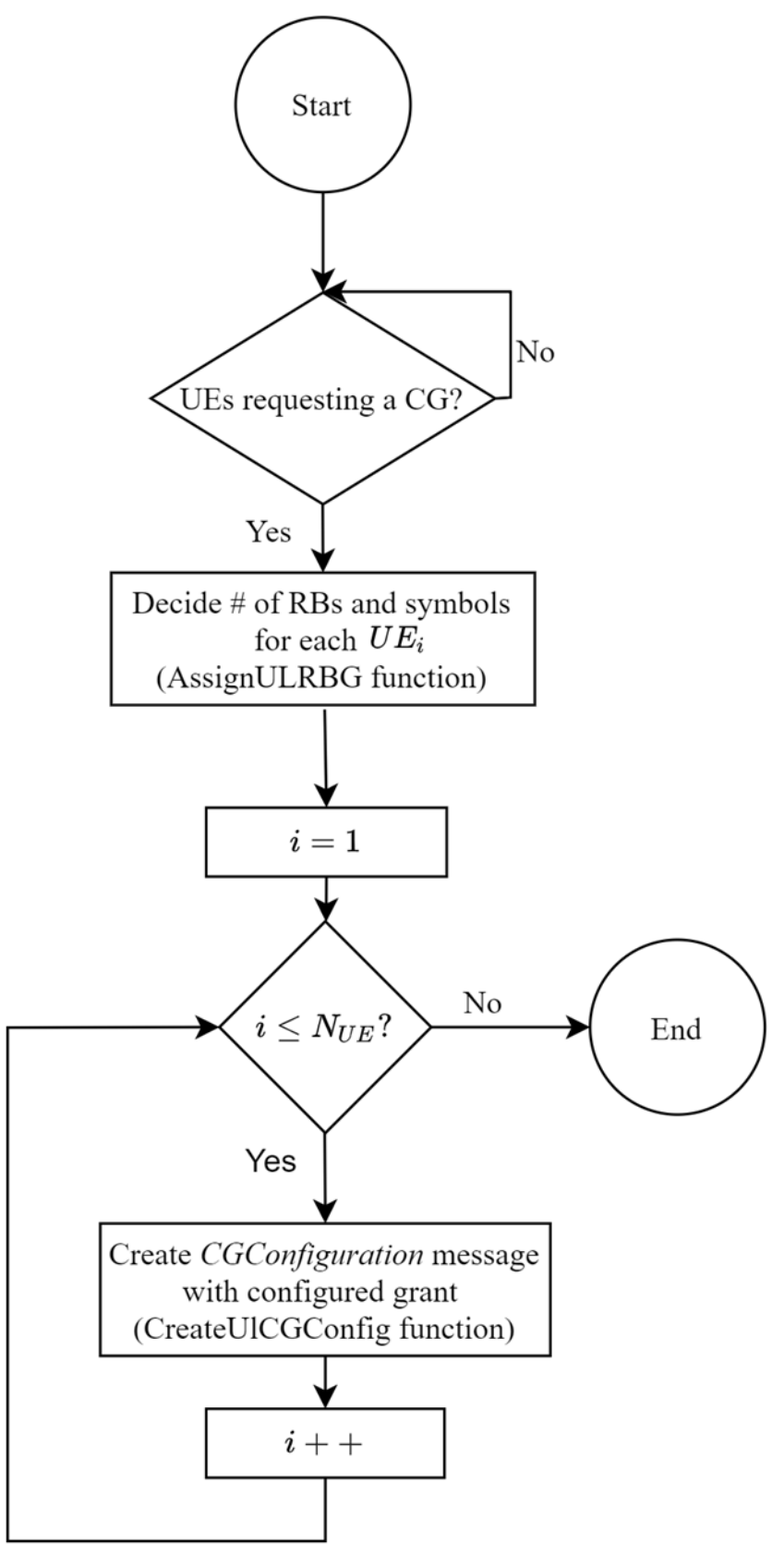

**Figure 5.** Radio resource allocation process for UL transmissions (DoScheduleUlData function) in 5G-LENA.

### 4.2. Flexible multiple access and radio resource allocation

Flexibility in radio resource allocation is constrained by the multiple access scheme used in the system. OFDMA in 5G NR allows the scheduling of UEs in the time and frequency domain. However, none of the multiple access schemes implemented in 5G-LENA implements the real capabilities of OFDMA in 5G NR. The 5GL-TDMA, 5GL-OFDMA, and 5GL-OFDMA with variable TTI multiple access schemes assume different constraints for scheduling radio resources in frequency or time among different UEs (see Section III). The constraints introduced by the multiple access schemes implemented in 5G-LENA may lead to inefficient use of radio resources. UEs may receive a higher number of radio resources than they actually need. This is especially the case when traffic is characterized by small packet sizes. This can result in higher latencies that can be detrimental to time-critical services.

We have extended the 5G-LENA MAC layer to implement a more accurate version of OFDMA in 5G NR. The objective is to provide an open-source software tool that accurately simulates the actual features and capabilities of 5G NR to support latency-critical services. The new OFDMA implemented in 5G-LENA allows allocating $r$ RBs in s consecutive OFDM symbols to a UE, with $r$ and $s$ being any integer number in [1, $R_{BW}$] and [1, $S_{slot}$], respectively ($s$ is always equal to $S_{slot}$ with 5GL-OFDMA, and $r$ is equal to $R_{BW}$ with 5GL-TDMA).

OFDMA has been implemented in the NrMacSchedulerOfdma class (see Figure 2). This class contains AssignULRBG and CreateUlCGConfig functions. DoScheduleUlData function in the MAC layer of the gNB in 5G-LENA (Figure 5) is responsible for calling these functions. As mentioned in the previous subsection, AssignULRBG function decides the number of OFDM symbols and RBs that should be allocated to each UE. The number of OFDM symbols and RBs allocated to each UE depends on several factors. First, it depends on the amount of data to be transmitted by the UE and the MCS to use in the packet transmission. The decision also depends on the multiple access scheme used. In this context, the AssignULRBG function included in 5G-LENA has been modified to allow the allocation of radio resources among different UEs simultaneously in frequency and time according to OFDMA in 5G NR. The final radio resource allocation depends on the scheduling policy implemented.

Finally, we have also modified the order followed in 5G-LENA to allocate radio resources for UL transmissions within a slot. As presented in section III, radio resources are allocated from the last to the first symbol within a slot in 5G-LENA. This order is not established in 5G NR. Therefore, we have eliminated this condition. Resources can now be allocated from the first to the last symbol within a slot, which is also key to guarantee very low latency.

### 4.3. Scheduling policies

Two new scheduling policies have been implemented in 5G-LENA to be applied with CG. These scheduling policies exploit the flexibility of OFDMA in 5G NR. The designed scheduling policies decide the number $r_i$ of RBs and the number $s_i$ of OFDM symbols allocated to each UE$_i$, with $i \in \mathbb{N}$ and $i \in [1, N_{UE}]$. Both scheduling schemes aim to use radio resources efficiently and minimize the experienced latency. To this end, they try to allocate the lowest number of radio resources that satisfies the radio resource demand of each UE. The defined scheduling policies aim to show the importance of accurately emulating the capabilities and flexibility of 5G NR, and they do not search for optimal solutions. This flexibility can be exploited to increase performance and more efficiently support latency-critical services.

The first scheduling policy minimizes the number of OFDM symbols allocated to each UE, referred to as Sym-OFDMA scheduler. The operation of Sym-OFDMA is presented in Algorithm I. Sym-OFDMA serves UEs following a first-come, first-served basis, i.e., from UE$_1$ to UE$_{N_{UE}}$ (line 3 in Algorithm I). Sym-OFDMA allocates radio resources from the first to the last symbol within a slot (variables $n_{slot}$, $n_s$, and $n_{RB}$ represent the slot, symbol within the slot, and RB, respectively, that is currently being allocated to a UE). Each UE$_i$ demands $d_i$ radio resources that is calculated as a function of the size of the packet to transmit and the MCS to use in the packet transmission. If the number $d_i$ of radio resources demanded by a UE$_i$ is equal to or lower than $R_{BW}$ (the number of RBs in a symbol), UE$_i$ receives $d_i$ consecutive RBs in an OFDM symbol, i.e., $r_i = d_i$ and $s_i = 1$ (see lines 4-5 in Algorithm I). If the UE$_i$ demands more than $R_{BW}$ radio resources ($d_i > R_{BW}$), UE$_i$ receives $R_{BW}$ RBs in $\lceil d_i / R_{BW} \rceil$ consecutive OFDM symbols (line 7 in Algorithm I). $r_{ini,i}$, $s_{ini,i}$ and $slot_i$ represent the first RB, OFDM symbol, and slot, respectively, allocated to UE$_i$ (see line 15). UE$_{i+1}$ will receive RBs in the same OFDM symbol as UE$_i$ if the number of unallocated RBs are enough to satisfy the demand of UE$_{i+1}$ (see lines 15-16). Otherwise, UE$_{i+1}$ will receive RBs in the next OFDM symbol (lines 9-11). Sym-OFDMA also considers that a UE can only receive RBs within a slot. In case there are not enough unallocated RBs and symbols to meet $d_i$ in the current slot, the UE$_i$ will be served in the next slot (lines 12-14). Figure 6.a shows an example of the radio resource allocation done by Sym-OFDMA in a scenario where 7 UEs demand 4 radio resources each to transmit using configured grant. Figure 6 represents the radio resources in one slot of 14 OFDM symbols, where the first and last symbols are reserved for control signals, and bandwidth is divided in $R_{BW}$=10 RBs.

ALGORITHM I: *SYM-OFDMA*

1. Input: $d_i \; \forall \; i \in [1, N_{UE}]$
2. $n_{slot}=1,\; n_s=1,\; n_{RB}=1$
3. **For** $i$=1 to $N_{UE}$
4. **If** $d_i \leq R_{BW}$
5. $r_i = d_i,\; s_i = 1$
6. **Else**
7. $r_i = R_{BW},\; s_i = \lceil d_i / R_{BW} \rceil$
8. **End If**
9. **If** $R_{BW} - n_{RB} + 1 < r_i$
10. $n_{RB} = 1,\; n_s = n_s + s_i$
11. **End If**
12. **If** $n_s + s_i - 1 > S_{slot}$
13. $n_{slot} = n_{slot} + 1,\; n_s = 1,\; n_{RB} = 1$
14. **End If**
15. $r_{ini,i} = n_{RB},\; s_{ini,i} = n_s,\; slot_i = n_{slot}$
16. $n_{RB} = n_{RB} + r_i$
17. **End For**

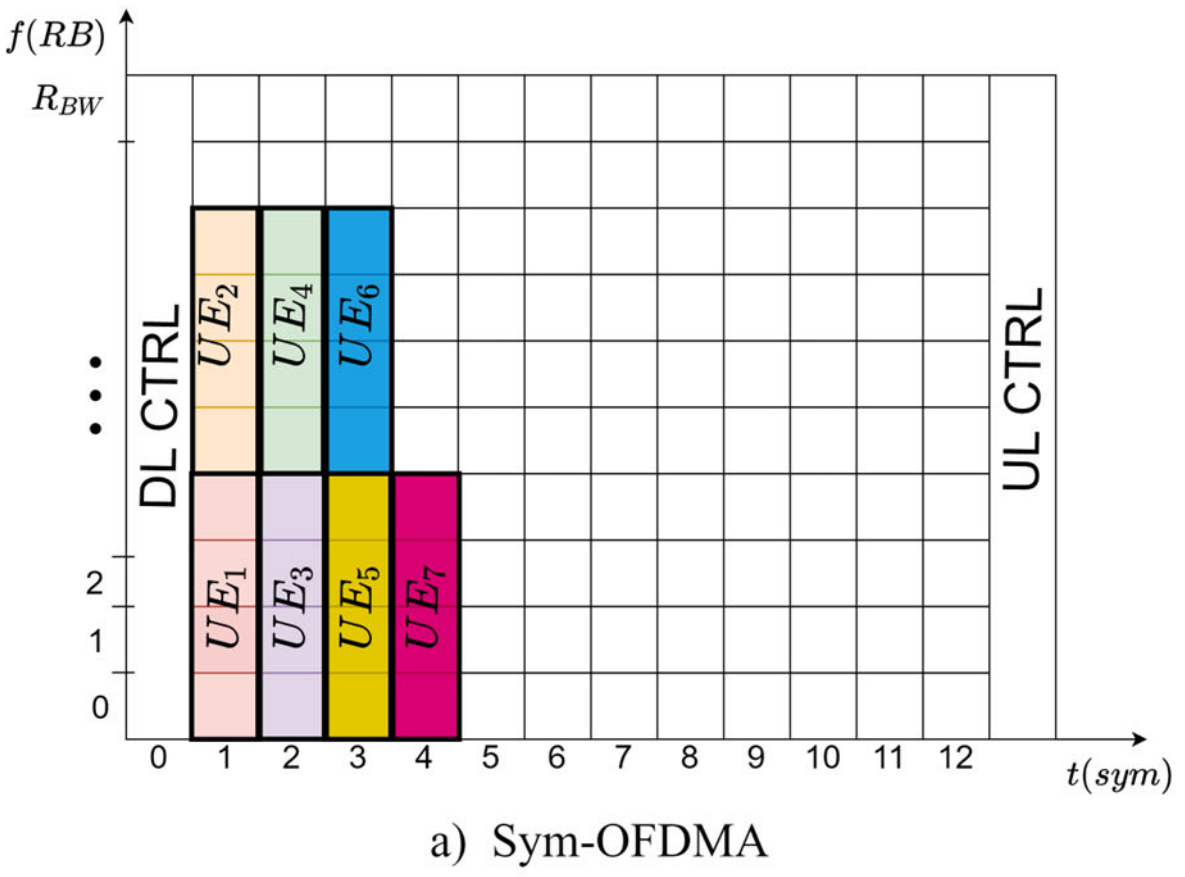

a) Sym-OFDMA

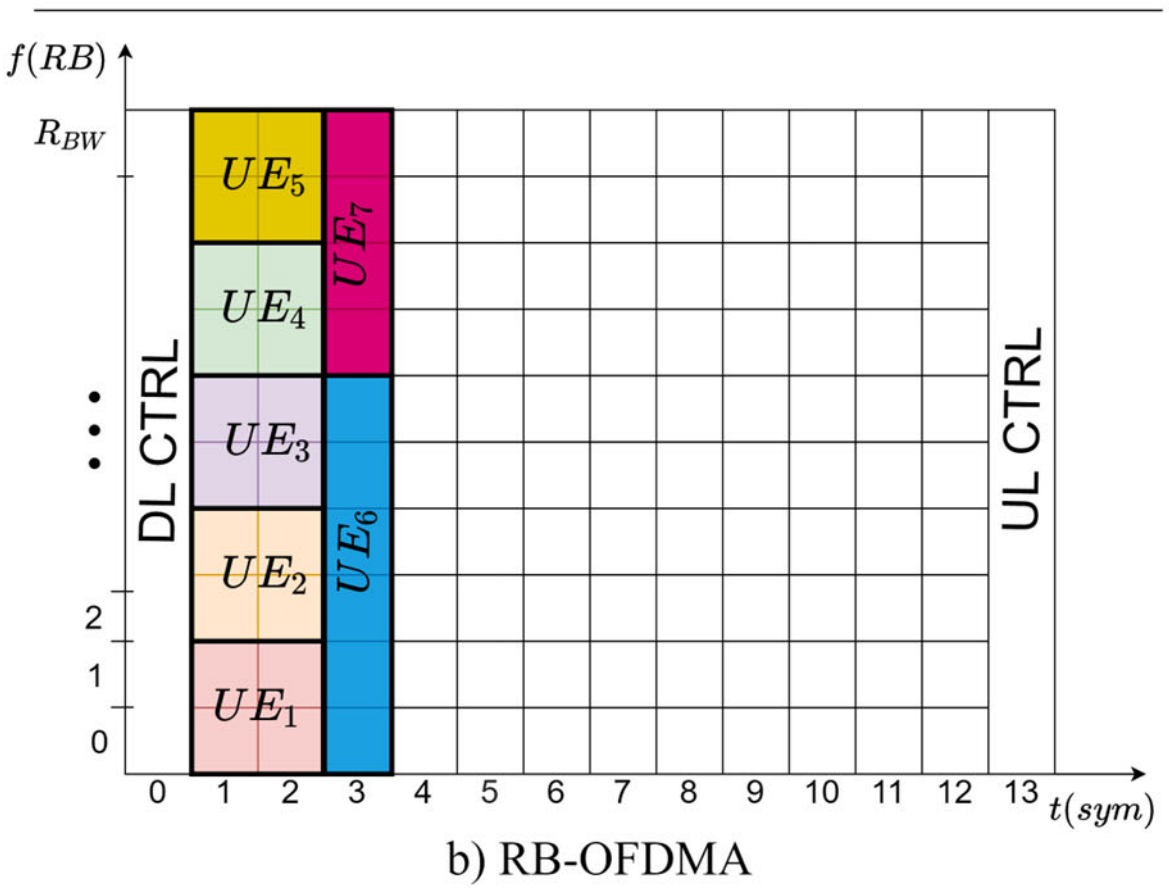

b) RB-OFDMA

**Figure 6**. Radio resource allocation with the designed scheduling policies.

The second scheduling policy calculates the number of radio resources ($r_i$) and OFDM symbols ($s_i$) that should be allocated to each $UE_i$ to minimize the number of RBs that are not allocated to UEs. It also establishes that each $UE_i$ has to receive at least $r_{min}$ RBs ($r_{min}$ can take any integer number[2] between 1 and $R_{BW}$). This second scheduling policy is referred to as RB-OFDMA. The operation of RB-OFDMA is presented in Algorithm II. RB-OFDMA also serves UEs following a first-come, first-served order, and allocates radio resources from the first to the last symbol within a slot. RB-OFDMA distributes the $R_{BW}$ RBs in a symbol among the maximum number of UEs ($n_{UE}$) considering that $r_i \geq r_{min}$ for each $UE_i$, i.e., $n_{UE}=\lfloor R_{BW}/r_{min} \rfloor$ (line 3 of Algorithm II). The UEs are divided in sets of $n_{UE}$ UEs (line 5); the last set of UEs can have less than $n_{UE}$. The UEs in each set $\phi$ will share the RBs in the same OFDM symbols. If the number of UEs in a set $\phi$ is equal to $n_{UE}$, each $UE_i$ in $\phi$ will receive $r_i = r_{min}$ RBs (lines 7 and 8). If the number of UEs in a set $\phi$ is lower than $n_{UE}$, the first $((n_{UE}/n)-\lfloor n_{UE}/n \rfloor)\cdot n$ UEs in $\phi$ will receive $r_i = \lceil n_{UE}/n \rceil \cdot r_{min}$ RBs (line 10). The rest of UEs in $\phi$ will receive $r_i = \lfloor n_{UE}/n \rfloor \cdot r_{min}$ RBs (line 11). Based on the number of RBs allocated to each UE, RB-OFDMA calculates the number $d_i^S$ of OFDM symbols that each $UE_i$ needs to meet its demand $d_i$ (line 14). Once $d_i^S$ is known, all UEs in a group $\phi$ receive the number of RBs necessary to satisfy the UE that requires a greater number of OFDM symbols (lines 16-17). RB-OFDMA also considers that a UE can only receive RBs within a slot. In case there are not enough unallocated RBs and symbols to meet $r_i$ and $s_i$ in the current slot, the $UE_i$ will be served in the next slot (lines 18-20). Figure 6.b shows an example of the radio resource allocation done with RB-OFDMA. The example considers that 7 UEs demand 4 radio resources to transmit their data using configured grant and $r_{min}$=2. In this example, the $N_{UE}$ UEs are divided into two sets of UEs. The first set includes 5 UEs (equal to $\lfloor R_{BW}/r_{min} \rfloor$) that receive $r_i$ =2 RBs and $s_i$=2 OFDM symbols. The second set includes 2 UEs that receive $r_i$ =4 RBs and $r_i$ =6 RBs for $UE_7$ and $UE_6$, respectively, and $s_i$=1 OFDM symbol.

ALGORITHM II: *RB-OFDMA*

1. Input: $d_i \ \forall \ i \in [1, N_{UE}]$, $r_{min}$
2. $n_{slot}=1$, $n_s=1$, $n_{RB}=1$, $n_{UEwithRB}=0$
3. Define $n_{UE}=\lfloor R_{BW}/r_{min} \rfloor$
4. **While** there are UEs without resources
5. Create set $\phi$ with $UE_i \ \forall i$ such that $i=[n_{UEwithRB}+1, \min(n_{UEwithRB}+n_{UE}, N_{UE})]$
6. Define $n$=number of UEs in $\phi$
7. **If** $n = n_{UE}$
8. $r_i = r_{min} \ \forall \ UE_i$ in $\phi$
9. **Else**
10. $r_i = \left\lceil \frac{n_{UE}}{n} \right\rceil \cdot r_{min}$ for the first $\left(\frac{n_{UE}}{n} - \left\lfloor \frac{n_{UE}}{n} \right\rfloor\right) \cdot n$ UEs in $\phi$
11. $r_i = \left\lfloor \frac{n_{UE}}{n} \right\rfloor \cdot r_{min}$ for the last $n-\left(\frac{n_{UE}}{n} - \left\lfloor \frac{n_{UE}}{n} \right\rfloor\right) \cdot n$ UEs in $\phi$
12. **End If**
13. **For** all $UE_i$ in $\phi$
14. $d_i^S = d_i / r_i$, $r_{ini,i} = n_{RB}$, $n_{RB} = n_{RB} + r_i$
15. **End For**
16. $d^S = \max_i \{d_i^S\} \ \forall \ UE_i$ in $\phi$
17. $s_i = d^S \ \forall \ UE_i$ in $\phi$
18. **If** $n_s = n_s + d^S - 1 > S_{slot}$
19. $n_{slot} = n_{slot}+1$, $n_s=1$
20. **End if**
21. $s_{ini,i}=n_s$, $slot_i = n_{slot} \ \forall \ UE_i$ in $\phi$
22. $n_s = n_s + d^S$, $n_{RB}=1$
23. $n_{UEwithRB} = n_{UEwithRB} + n_{UE}$
24. **End While**

The value of $r_{min}$ can be tuned to optimize radio resource efficiency. In this work, we calculate $r_{min}$ with the aim of minimizing the number of RBs not allocated to any UE. To this end, we search for the value of $r_{min}$ that minimizes the number of not allocated RBs in a symbol (lines 1-6 of Algorithm III). If several values satisfy this condition, we select the value that provides the lowest difference between the number of allocated radio resources to UEs ($r_i \cdot s_i = x \cdot d_i^S(x)$) and their demands (line 8).

ALGORITHM III: $r_{min}$ in *RB-OFDMA*

1. $aux=R_{BW}$
2. Create $X=\{x\}$ with $x \in \mathbb{N}$, $1<x<R_{BW}$, and $\mathrm{mod}(aux/x)=0$.
3. **If** $X$ is empty
4. $aux=aux-1$
5. Go to 3
6. **End If**
7. Calculate number $d_i^S(x)$ of OFDM symbols needed to meet the UE demands $d_i$ when UEs receive x RBs, $\forall i \in [1,N_{UE}]$ and $\forall x \in X$.
8. Set $r_{min} = x \in X$ that satisfies $\min_{i,x} (x \cdot d_i^S(x) - d_i)$

## 5. Evaluation Scenario & Reference Schemes

We consider an evaluation scenario where a single 5G NR cell covers a typical work cell of 10 x 10 $m^2$ where a closed-loop control application is implemented. The cell is assigned a bandwidth of $BW$ in the 3.7–3.8 GHz band[3], with $BW$ equal to 10, 20, or 40 MHz, and operates in TDD mode [21]. We consider slots with 14 OFDM symbols and evaluate 3 slot configurations for which the last 13, 9, and 5 OFDM symbols are used for UL and the rest of OFDM symbols for DL. We refer to the different configurations as 1D13U, 5D9U, and 9D5U, respectively. The first and last symbols within a slot are reserved for the transmission of the control channels in DL and UL, respectively. We consider the use of 30 kHz SCS as recommended in [22] for industrial environments (we also evaluate the use of 15 and 60 kHz when specifically indicated). There are $N_{UE}$=15 sensors randomly distributed in the work cell, and we consider that they are in the Line of Sight with the gNB. The radio channel is characterized by fast-fading and shadowing, and they follow a spatial channel model defined in [23]. Following [24], sensors generate periodic data packets that are transmitted to a central monitoring system. All sensors participating in the closed-loop control application generate data packets simultaneously. Data packets are characterized by a size $p_i$ that is set equal to 10 or 25 bytes and a periodicity equal to 10 ms [24]. Sensors transmit the data packets in UL to the gNB towards the central monitoring system. The IPv4 header with a value of 22 bytes is added to this packet size, $p_{headers}$. We also consider the use of cyclic redundancy check (CRC) code. All sensors use MCS 12 in MCS table 1 [12] for the periodic traffic transmissions; MCS 12 provides a good trade-off between robustness and transmission rate in the considered scenario. Furthermore, we consider the use of one MIMO transmission layer (referred to as $v$). Processing times in the UE and gNB are calculated as indicated in [25] and [12]. The values for the different evaluation parameters are summarized in Table 1[4].

[2] $r_{min}$ can be configured based on traffic characteristic to optimize system performance.
[3] The 3.7–3.8 GHz band is considered in some European countries for non-public network deployments [28].

**Table 1.** Evaluation parameters.

| Parameters | Value |
|---|---|
| Simulation Duration | 10 seconds |
| Work Cell dimensions | 10 x 10 m$^2$ |
| Number of UEs ($N_{UE}$) | 15 |
| Packet Size ($p_i$) | 10 or 25 bytes |
| Packet Periodicity | 10 ms |
| Frequency Band | 3.7 – 3.8 GHz |
| Bandwidth (BW) | 10, 20 or 40MHz |
| Numerology ($\mu$) | 1 (SCS = 30 kHz) |
| MCS | 12 |
| MIMO transmission layers ($v$). | 1 |

We compare the performance achieved with the designed CG scheduling policies when 5G NR OFDMA is used with the performance obtained using the 5GL-TDMA and 5GL-OFDMA schemes already implemented in 5G-LENA. To make a fair comparison, a First Come First Served scheduler is applied with 5GL-TDMA and 5GL-OFDMA. In this context, the scheduler serves UEs following a first-come, first-served basis, i.e., from $UE_1$ to $UE_{N_{UE}}$. Each UE receives the required number of RBs and OFDM symbols to satisfy its demand. When 5GL-TDMA is used, all UEs will receive $r_i = R_{BW}$, and $s_i$ will be calculated for each UE as $s_i=\lceil d_i/R_{BW}\rceil$ to meet its demand. When 5GL-OFDMA is applied, all UEs will receive $s_i= S_{slot}$ OFDM symbols, and $r_i$ is calculated as $r_i=\lceil d_i/S_{slot}\rceil$.

## 6. Analytical Validation

In this section, we derive the analytical expressions that model the maximum UL latency experienced by the UEs using CG and the different scheduling policies and multiple access schemes. In this work, the UL latency accounts for the elapsed time from when a packet is created at the RLC layer of a UE until it is received at the RLC layer of the gNB. We compare the results achieved analytically with those achieved by simulation in order to validate the implementation in CG in 5G-LENA.

### 6.1. Analytical modeling

This subsection derives the analytical expressions of the maximum UL latency ($L_{UL}$) experienced by the UEs. We consider that all UEs transmit packets of the same size. Following [26], the latency of a packet transmitted in UL with CG is given by the following components. First, we need to account for the processing times at the UE and the gNB that represent the time required to generate the packet at the transmitter and decode the data at the receiver, respectively ($t_{UE,tx}$, and $t_{gNB,rx}$). Other latency components are the frame alignment time ($t_{fa}$) that accounts for the time interval from the creation of a packet until the next transmission opportunity for the Physical Uplink Shared Channel (PUSCH), the waiting time for the resources allocated for the packet transmission ($t_w$), and the transmission time of the packet ($t_{tt}$). $L_{UL}$ is then calculated as:

$$L_{UL}=t_{UE,tx}+t_{fa}+ t_w+ t_{tt} + t_{gNB,rx} \quad (1)$$

The processing time at the transmitter and the receiver ($t_{UE,tx}$ and $t_{gNB,rx,}$) are calculated as indicated in [23]. $t_{fa}$ depends on the frame structure, i.e., on the configuration of slots and OFDM symbols within a slot for the transmission of control and data channels for DL and UL transmissions. The waiting time $t_w$ accounts for the time interval that a UE has to wait after $t_{fa}$ for the assigned resources. $t_w$ depends on the way radio resources are allocated to the UEs. Therefore, $t_w$ depends on the multiple access mode and the scheduling policy. The transmission time $t_{tt}$ is equal to the time duration of the OFDM symbols used for the packet transmission, and it can be calculated as $s_i{\cdot}T_{sym}$, $T_{sym}$ is the duration of an OFDM symbol. $T_{sym}$ depends on the numerology used for the transmission of the packet, and $s_i$ depends on the multiple access mode and scheduling scheme. For numerology 1 with

[4] We ran sufficient simulations to achieve statistically valid results: a minimum of 100 runs (each run simulates 10 s of network operation) for each evaluated configuration was executed.

SCS 30 kHz, $T_{sym}$ is equal to 35.67 ms. We calculate $t_{fa}$, $t_{w,}$ and $t_{tt}$ for the different scheduling policies and multiple access schemes studied in this work.

#### 6.1.1. 5GL-TDMA

5GL-TDMA allocates all RBs in an OFDM symbol to the same UE. As a result, a UE$_i$ receives $r_i^{\text{5GL-TDMA}}=R_{BW}$ RBs and $s_i^{\text{5GL-TDMA}}=\lceil d_i/R_{BW}\rceil$ OFDM symbols. $R_{BW}$ depends on the bandwidth and is given in [21].

$t_{tt}$ for a UE$_i$ can be calculated as $\lceil d_i/R_{BW}\rceil\cdot T_{sym}$, where $d_i$ is the number of radio resources demanded by UE$_i$. $d_i$ can be calculated as a function of the packet size ($p_i$) and the MCS used to transmit the packet as:

$$d_i=\left\lceil\frac{(tbs_i(p_i+p_{headers})+\text{CRC})}{R\cdot Q_m\cdot v\cdot N_{sc,RB}}\right\rceil \tag{2}$$

In (2), $tbs_i(p_i+p_{headers})$ represents the smallest transport block size from the available values given in [27] that can be used to transmit $p_i+p_{headers}$ bits. $Q_m$ is the modulation order, $R$ is the code rate, and $v$ is the number of MIMO transmission layers. The waiting time for the last UE served (the one that experiences the highest latency) is equal to the transmission time required by the other $N_{UE}$-1 UEs requesting resources. Therefore, we can calculate $t_w+t_{tt}$ as:

$$t_w+t_{tt}=\sum_{i=1}^{N_{UE}}\left\lceil\frac{d_i}{R_{BW}}\right\rceil\cdot T_{sym} \tag{3}$$

If all UEs transmit packets of the same size, (3) can be expressed as:

$$t_w+t_{tt}=N_{UE}\cdot\left\lceil\frac{d_i}{N_{RB}}\right\rceil\cdot T_{sym} \tag{4}$$

$t_{fa}$ accounts for the delay introduced by the transmission of other channels (control channels in UL and DL -PUCCH or PDCCH- or the transmission of the PDSCH). We consider that the first and last OFDM symbols of a slot are reserved for the transmission of control channels and 12 ODFM symbols are used for data transmission in UL. To calculate $t_{fa}$, we need to know how many slots are needed to allocate resources for all UEs, which is represented as $n_{slot}$, and is calculated as:

$$n_{slot}=\left\lceil\frac{\sum_{i=1}^{N_{UE}} s_i^{5GL\text{-}TDMA}}{12}\right\rceil=\left\lceil\frac{N_{UE}\cdot s_i^{5GL\text{-}TDMA}}{12}\right\rceil \tag{5}$$

The first and last OFDM symbols in each slot are dedicated to control channels transmissions. Then, $t_{fa}$ is calculated as:

$$t_{fa}=(2\cdot n_{slot}-1)\cdot T_{sym} \tag{6}$$

Using (4), (5), and (6), the maximum latency experienced by UEs using 5GL-TDMA is calculated as:

$$L_{UL}=t_{UE,tx}+(2\cdot n_{slot}-1+N_{UE}\cdot\left\lceil\frac{d_i}{R_{BW}}\right\rceil)\cdot T_{sym}+t_{gNB,rx} \tag{7}$$

#### 6.1.2. 5GL-OFDMA

5GL-OFDMA allocates the same RB in all OFDM symbols within a slot to the same UE. 5GL-OFDMA then allocates $r_i^{\text{5GL-OFDMA}}=\lceil d_i/S_{slot}\rceil$ RBs in $s_i^{\text{5GL-OFDMA}}=S_{slot}$ OFDM symbols for each UE$_i$ ($d_i$ is calculated in (2)). The latency $t_{fa}+t_w+t_{tt}$ for 5GL-OFDMA is calculated as the time needed to transmit $n_{slot}$ slots minus the time duration of the last OFDM symbol. $n_{slot}$ is the number of slots used to transmit the packets of all UEs. $t_{fa}+t_w+t_{tt}$ is then calculated as:

$$t_{fa}+t_w+t_{tt}=n_{slot}\cdot T_{slot}-T_{sym}=(14\cdot n_{slot}-1)\cdot T_{sym} \tag{8}$$

In (8), $T_{slot}$ is the time duration of a slot, which is equal to $14\cdot T_{sym.}$ The parameter $n_{slot}$ in (8) is calculated as:

$$n_{slot}=\left\lceil\sum_{i=1}^{N_{UE}} r_i^{\text{5GL-OFDMA}}/R_{BW}\right\rceil=\left\lceil\frac{N_{UE}\cdot r_i^{\text{5GL-OFDMA}}}{R_{BW}}\right\rceil \tag{9}$$

Using (9) and (10), the maximum latency experienced by UEs using 5GL-OFDMA is calculated as:

$$L_{UL}=t_{UE,tx}+(14\cdot n_{slot}-1)\cdot T_{sym}+t_{gNB,rx} \tag{10}$$

#### 6.1.3. Sym-OFDMA

Sym-OFDMA allocates to each UE$_i$ the number of RBs necessary to satisfy $d_i$ in the minimum number of OFDM symbols possible. We can then distinguish two cases. When $d_i > R_{BW}$, the UE$_i$ receives $r_i^{Sym\text{-}OFDMA}=R_{BW}$ RBs in $s_i^{Sym\text{-}OFDMA}=\lceil d_i/R_{BW}\rceil$ consecutive symbols. When $d_i\le R_{BW}$, Sym-OFDMA allocates $r_i^{Sym\text{-}OFDMA}=d_i$ RBs and $s_i^{Sym\text{-}OFDMA}=1$ to UE$_i$. In this case, several UEs can share the RBs in the same OFDM symbol. The number of UEs sharing RBs in the same OFDM symbol is given by $\lfloor R_{BW}/d_i\rfloor$ when all UEs demand the same number $d_i$ of RBs. We can then calculate the number of OFDM symbols $n_s$ needed to serve all UEs as:

$$n_s=\begin{cases}\dfrac{N_{UE}}{\lfloor R_{BW}/d_i\rfloor}, & d_i\leq R_{BW}\\ \left\lceil\dfrac{d_i}{R_{BW}}\right\rceil\cdot N_{UE}, & d_i>R_{BW}\end{cases} \quad (11)$$

Using (11), the number of slots needed to serve all UEs is equal to $n_{slot}=\lfloor n_s/12\rfloor$; each slot has 12 OFDM symbols for UL data transmission. Considering that the first and last OFDM symbols within slots are used for control channels, $t_{fa}+t_w+t_{tt}$ can be calculated as:

$$t_{fa}+t_w+t_{tt}=(2\cdot n_{slot}-1+n_{sym})\cdot T_{sym} \quad (12)$$

The maximum latency experienced using Sym-OFDMA is calculated as:

$$L_{UL}=t_{UE,tx}+\left(2\cdot n_{slot}-1+n_{sym}\right)\cdot T_{sym}+t_{gNB,rx} \quad (13)$$

6.1.4. RB-OFDMA

RB-OFDMA establishes that each $UE_i$ has to receive at least $r_{min}$ RBs. Considering this constraint, it distributes the $R_{BW}$ RBs in a symbol among the maximum number of UEs. The maximum number of UEs that can share the RBs in an OFDM symbol is given by $n_{UE,1}=\lfloor R_{BW}/r_{min}\rfloor$. In this context, the $N_{UE}$ UEs are divided in $\lfloor N_{UE}/n_{UE,1}\rfloor$ groups of $n_{UE,1}$ UEs that receive $r_i^{RB\text{-}OFDMA(1)}=r_{min}$ RBs in $s_i^{RB\text{-}OFDMA(1)}=\lceil d_i/r_{min}\rceil$ OFDM symbols, and one last group with $n_{UE,2}=N_{UE}-\lfloor N_{UE}/n_{UE,1}\rfloor\cdot n_{UE,1}$ UEs that receive $r_i^{RB\text{-}OFDMA(2)}=\left\lfloor\lfloor R_{BW}/r_{min}\rfloor/n_{UE,2}\right\rfloor\cdot r_{min}$ RBs in $s_i^{RB\text{-}OFDMA(2)}=\lceil d_i/r_i^{RB\text{-}OFDMA(2)}\rceil$ OFDM symbols. We can then calculate the number $n_s$ of OFDM symbols necessary to serve all UEs as:

$$n_s=\lfloor N_{UE}/n_{UE,1}\rfloor\cdot s_i^{RB\text{-}OFDMA(1)}+s_i^{RB\text{-}OFDMA(2)} \quad (14)$$

Using (14), the number of slots needed to serve all UEs is calculated as $n_{slot}=\lfloor n_s/12\rfloor$. Considering that the first and last OFDM symbols within slots are used for control channels, $t_{fa}+t_w+t_{tt}$ can be calculated using (12) as $(2\cdot n_{slot}-1+n_{sym})\cdot T_{sym}$. The maximum latency experienced using RB-OFDMA is calculated as:

$$L_{UL}=t_{UE,tx}+t_{gNB,rx}+\left(2\cdot n_{slot}-1+\left\lfloor\frac{N_{UE}}{n_{UE,1}}\right\rfloor\cdot s_i^{RB\text{-}OFDMA(1)}+s_i^{RB\text{-}OFDMA(2)}\right)\cdot T_{sym}. \quad (15)$$

6.2. Validation

Figure 7 compares the maximum latency experienced in UL transmissions obtained analytically and through simulations when using CG with 5GL-TDMA, 5GL-OFDMA, Sym-OFDMA, and RB-OFDMA for different bandwidth values (*BW*) when packets size is equal to 10 bytes and slot format 1D13U is used. Figure 7 clearly shows that the simulated and analytical results precisely match all the multiple access schemes and scheduling policies evaluated. It is important to note that the maximum latency obtained with Sym-OFDMA and RB-OFDMA is equal to or lower than the maximum latency experienced with 5GL-TDMA and 5GL-OFDMA for all the evaluated *BW* values. The same maximum latency is achieved for all *BW*s when 5GL-TDMA is used. This is because 5GL-TDMA assigns all the RBs in an OFDM symbol to the same UE regardless of the number of RBs available or *BW* and its RB demand. As a result, 5GL-TDMA uses the same number of OFDM symbols to serve the $N_{UE}$ UEs for all the evaluated *BW* values, and the latency experienced by the last served UE does not change with *BW*. When 5GL-OFDMA is used, the maximum latency experienced by a UE decreases 43.1% when *BW* increases from 10 to 20 MHz. This is because two slots are needed to allocate resources to all the UEs when *BW*=10 MHz. When *BW* increases to 20 MHz, the radio resource demand of all UEs can be satisfied with the RBs in one slot. Since each UE receives RBs in all the OFDM symbols dedicated for data within a slot, the maximum latency experienced by a UE remains constant as *BW* increases above 20 MHz. Both Sym-OFDMA and RB-OFDMA provide the lowest maximum latency values for all the *BW* evaluated. In addition, the maximum latency experienced with both scheduling policies decreases when *BW* increases. This is due to the greater flexibility introduced by OFDMA that allows radio resources to be allocated more efficiently and the experienced latency to be reduced. The performance achieved with both Sym-OFDMA and RB-OFDMA is analyzed in more depth in the following section.

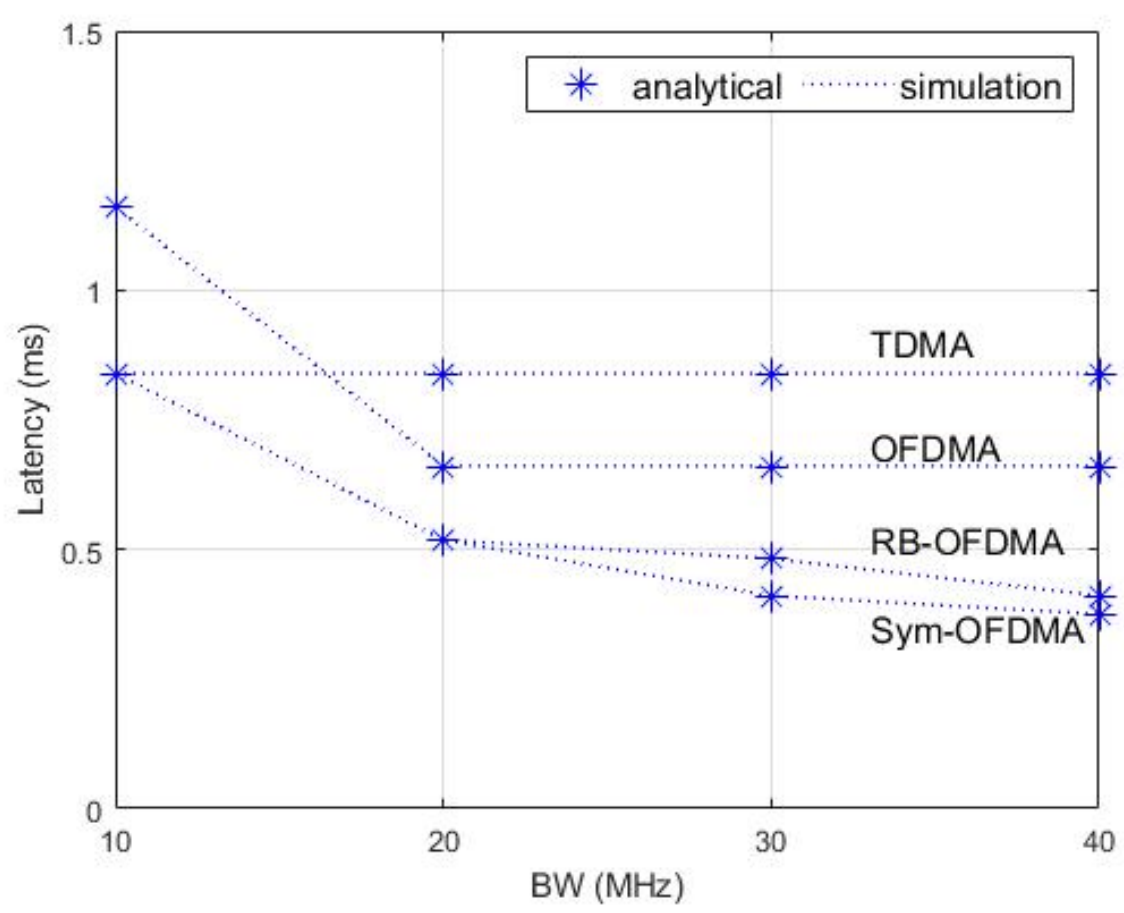

**Figure 7.** Maximum UL latency as a function of the bandwidth (packet size of 10 bytes).

We also compare the latency results obtained with CG using 5G-LENA with the analytical values reported in [25] (third column in Table 2). Table 2 also includes the latency results obtained using dynamic scheduling for performance comparison. We consider a scenario with only one UE that transmits empty packets using 2 OFDM symbols. We evaluate the use of numerologies ($\mu$) 0, 1, and 2, which correspond to a subcarrier spacing of 15, 30, and 60 kHz and a symbol duration of 71.4 µs, 35.6 µs, and 17.9 µs, respectively. We should note that the higher the numerology, the lower the symbol duration and, therefore, lower latency. We consider a slot format 1D13U and $BW$=20 MHz. The results in Table 2 show that the latency results obtained with CG implemented in 5G-LENA are in line with the analytical results presented in 3GPP TR 37.910 [25], which validates the implemented CG in 5G-LENA. Moreover, the latency obtained with configured grant has a reduction of 93.3% compared to the latency obtained with the dynamic scheduler in the case of numerology 0.

**Table 2**. Latency experienced with Dynamic Scheduling and Configured Grant

| $\mu$ | Dynamic scheduling | Configured Grant | |
|---|---|---|---|
| | | 5G-LENA | TR 37.910 [25] |
| 0 | 7.31 ms | 0.49 ms | 0.52 ms |
| 1 | 3.70 ms | 0.30 ms | 0.30 ms |
| 2 | 1.90 ms | 0.25 ms | 0.24 ms |

## 7. Performance evaluation

In this section, we analyze the performance achieved with the different multiple access schemes and scheduling policies in the evaluation scenario presented in section V.

### 7.1. Impact of the cell bandwidth

We analyze the latency that can be achieved with the different multiple access schemes and scheduling policies when different cell bandwidths are considered. Figure 8 shows a boxplot of the UL latency experienced by UEs using CG with 5GL-TDMA, 5GL-OFDMA, Sym-OFDMA, and RB-OFDMA when bandwidth is equal to 10, 20 and 40 MHz and packet size is 10 bytes. In Figure 8, the red line within the box represents the average of the experienced latency, and the edges of the box are the 10th and 90th percentiles. The crosses represent the minimum and maximum values. At least otherwise indicated, we consider a packet size of 10 bytes and slot format 1D13U. Figure 9 depicts the number of radio resources used with the different multiple access modes and scheduling policies calculated with respect to the number of radio resources used by 5GL-TDMA. Figure 8 shows that Sym-OFDMA and RB-OFDMA equal or reduce the maximum and average latency experienced by the UEs for all the evaluated $BW$ values compared with 5GL-OFDMA and 5GL-TDMA. This is due to a more efficient use of radio resources, as shown in Figure 9. When the bandwidth is low ($BW$=10 MHz), both Sym-OFDMA and RB-OFDMA achieve similar latency results to that achieved with 5GL-TDMA. 5GL-TDMA assigns all the RBs in a symbol to the same UE. In this case, the number of RBs needed to transmit 10-bytes packets is equal to the number of RBs available in the 10 MHz bandwidth. Sym-OFDMA then allocates all the RBs in a symbol to a UE, achieving the same latency performance as 5GL-TDMA. Figure 9.a also shows that both 5GL-TDMA and Sym-OFDMA use the same amount of radio resources. When BW=10MHz, RB-OFDMA allocates RBs in two consecutive symbols to each UE, and the RBs in a symbol are shared by two UEs. As a result, the maximum latency experienced with RB-OFDMA is the same as with 5GL-TDMA and Sym-OFDMA, as shown in Figure 8. Figure 9.a shows that RB-OFDMA also uses the same number of radio resources as 5GL-TDMA and Sym-OFDMA. 5GL-OFDMA provides the largest latency values when the bandwidth is 10 MHz. 5GL-OFDMA allocates RBs in all symbols within a slot to UEs. When bandwidth is equal to 10 MHz, only 8 of the 15 UEs can receive resources in the first slot after the packets are generated, and they experience a latency of 0.65 ms. The rest of the UEs receive resources in the next slot and experience a latency of 1.15 ms. Figure 9.a shows that 5GL-OFDMA uses 40% more radio resources than 5GL-TDMA, Sym-OFDMA and RB-OFDMA.

When the bandwidth increases from 20 to 40 MHz, 5GL-OFDMA reduces the maximum latency experienced compared with 5GL-TDMA. Figure 8 shows that all UEs experience the same latency when $BW$ is equal to 20 and 40 MHz. This happens because all UEs share RBs in all the OFDM symbols of the first slot after the packets are generated. Sym-OFDMA and RB-OFDMA achieve the lowest latency for the UEs. Furthermore, both Sym-OFDMA and RB-OFDMA use a lower number of radio resources to serve the $N_{UE}$ UEs than 5GL-TDMA and 5GL-OFDMA (Figure 9). Sym-OFDMA allows several UEs to allocate RBs in the same OFDM symbol. As $BW$ and the number of available RBs increase, more UEs receive RBs in the same

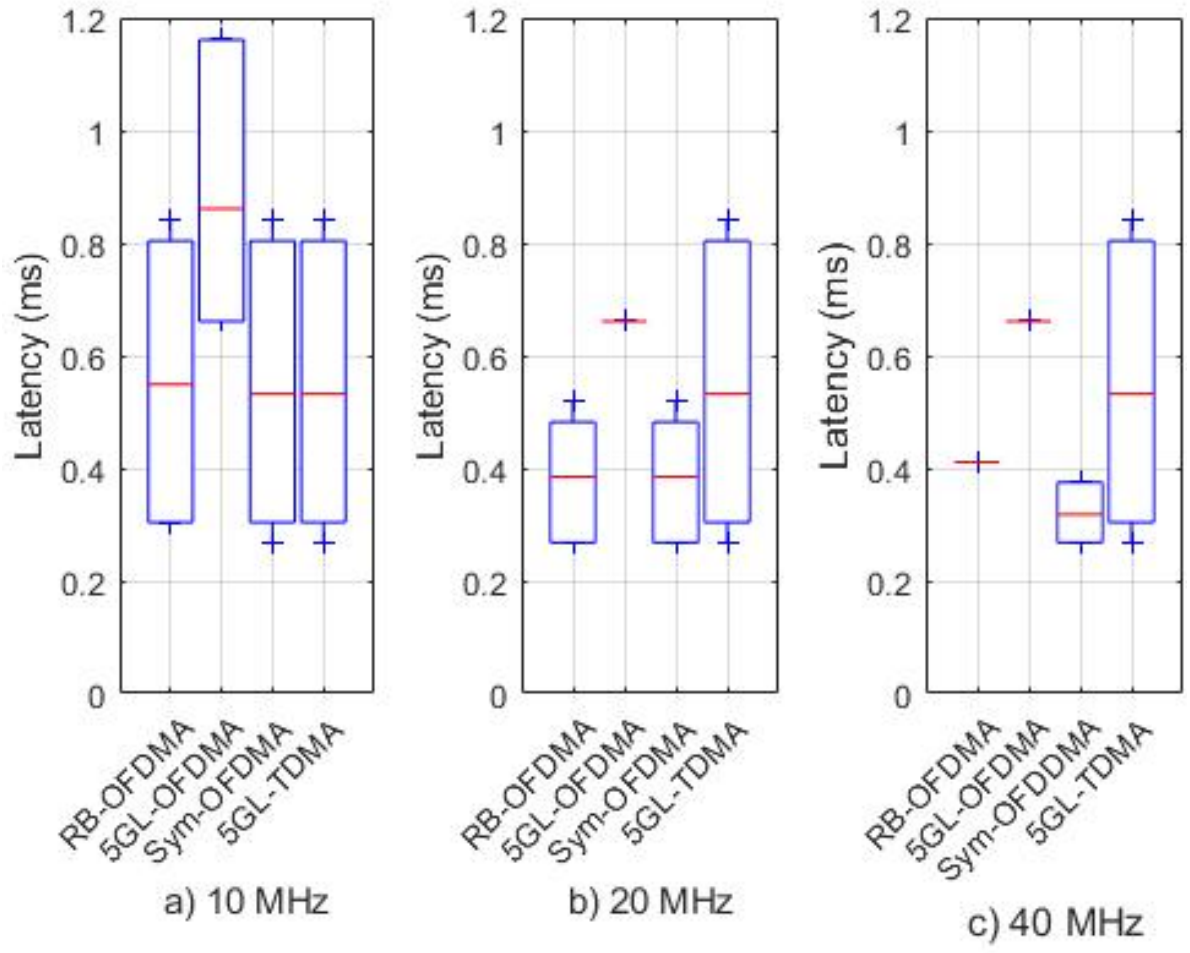

**Figure 8.** UL latency experienced as a function of the bandwidth (packet size of 10 bytes).

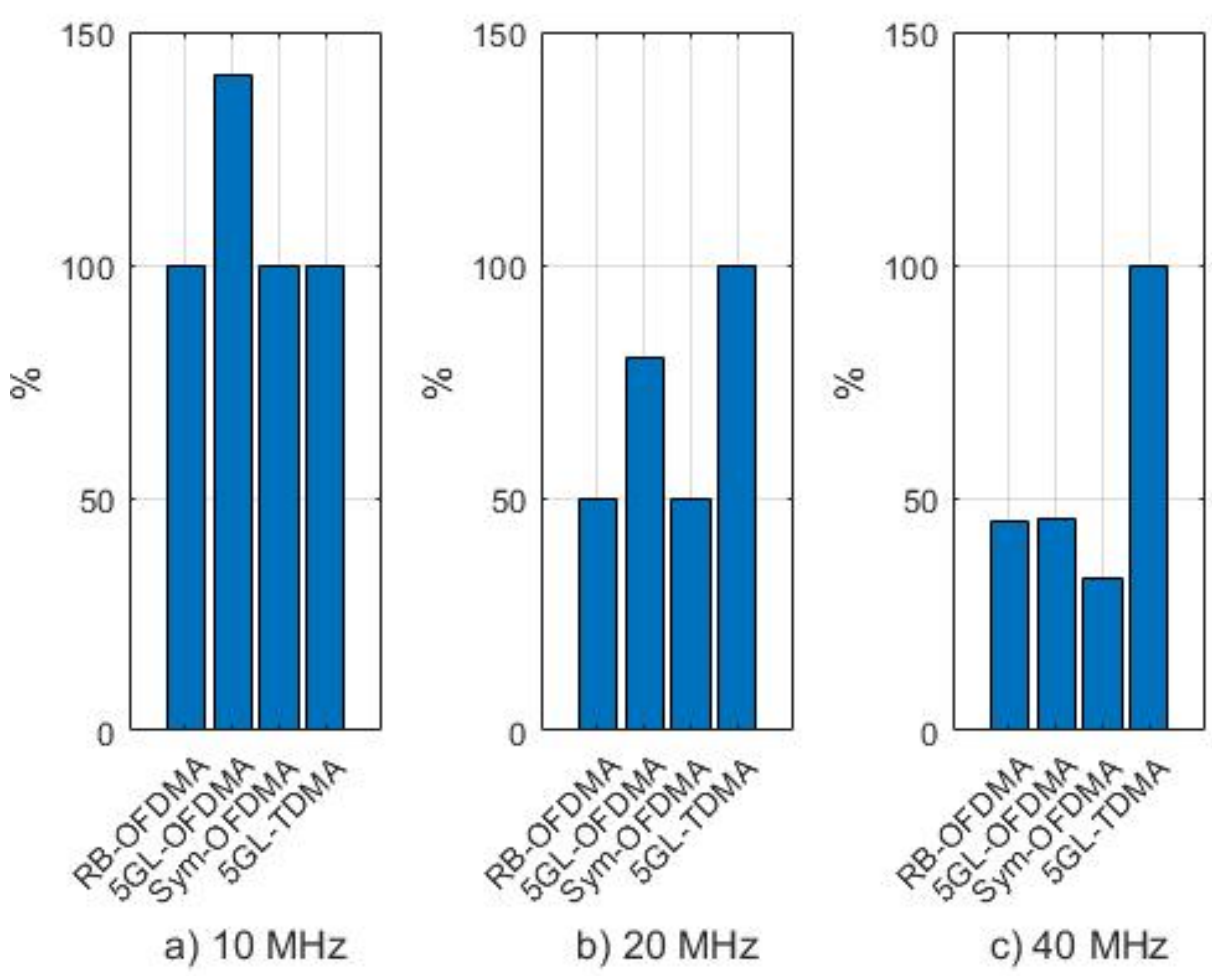

**Figure 9.** Percentage of radio resources used for the transmission of packets with respect to 5GL-TDMA (packet size of 10 bytes).

OFDM symbol. Sym-OFDMA then requires a lower number of OFDM symbols to serve all UEs compared with 5GL-TDMA. As a result, the latency experienced by the UEs decreases. Figure 8 also shows that the maximum latency experienced with RB-OFDMA also reduces when the bandwidth increases. However, the minimum and average latency experienced increases when bandwidth increases from 20 to 40 MHz. This is due to the different number of OFDM symbols allocated to the UEs when *BW* is equal to 20 and 40 MHz, respectively. When *BW*=20 MHz, each UE receives RBs in only one OFDM symbol, and the RBs in a symbol are shared by 2 UEs. When bandwidth is equal to 40 MHz, RB-OFDMA allocates RBs in 5 consecutive OFDM symbols to each UE, and the RBs in a symbol are allocated to all the UEs.

## 7.2. Impact of the packet size

Figure 10 shows the boxplot of the UL latency experienced by the UEs using CG with 5GL-TDMA, 5GL-OFDMA, Sym-OFDMA and RB-OFDMA when packets of 10 and 25 bytes are transmitted, respectively, and considering *BW*=20 MHz, and slot format 1D13U. Figure 10 shows that the latency experienced by the UEs increases with the packet size for Sym-OFDMA and RB-OFDMA, while it remains constant for 5GL-TDMA and 5GL-OFDMA. This is the case because 5GL-TDMA and 5GL-OFDMA allocate more radio resources than demanded by each UE when 10 bytes of data are transmitted per packet. The number of allocated radio resources is enough to satisfy the radio resource demand when the packet size increases to 25 bytes. Therefore, the experienced UL latency remains constant with 5GL-TDMA and 5GL-OFDMA because they allocate the same radio resources to UEs when the packet size is 10 and 25 bytes. On the other hand, Sym-OFDMA and RB-OFDMA allocate to each UE a number of radio resources more adjusted to their demands by taking full advantage of the 5G NR OFDMA flexibility. For that reason, the number of allocated radio resources to each UE increases when packet size increases. This also results in the increase of the UL latency experienced by the UEs. However, it is important to highlight that the maximum UL latency experienced with Sym-OFDMA and RB-OFDMA is always lower than the one experienced with 5GL-TDMA and 5GL-OFDMA, respectively. In the case of Sym-OFDMA, the number of available RBs is not enough to satisfy the demand of more than one UE, and each UE receives RBs in different OFDM symbols. As a result, Sym-OFDMA and 5GL-TDMA achieve the same latency performance. However, Sym-OFDMA only uses 66% of the radio resources used by 5GL-TDMA, and the non-allocated radio resources could be used by other UEs. RB-OFDMA has higher flexibility than Sym-OFDMA since it can allocate any number of RBs and OFDM symbols to each UE; the number of OFDM symbols is always 1 with Sym-OFDMA if the radio resource demand of the UE can be satisfied with the available RBs in an OFDMA symbol. Thanks to this, it achieves a lower maximum UL latency when the packet size is equal to 25 bytes.

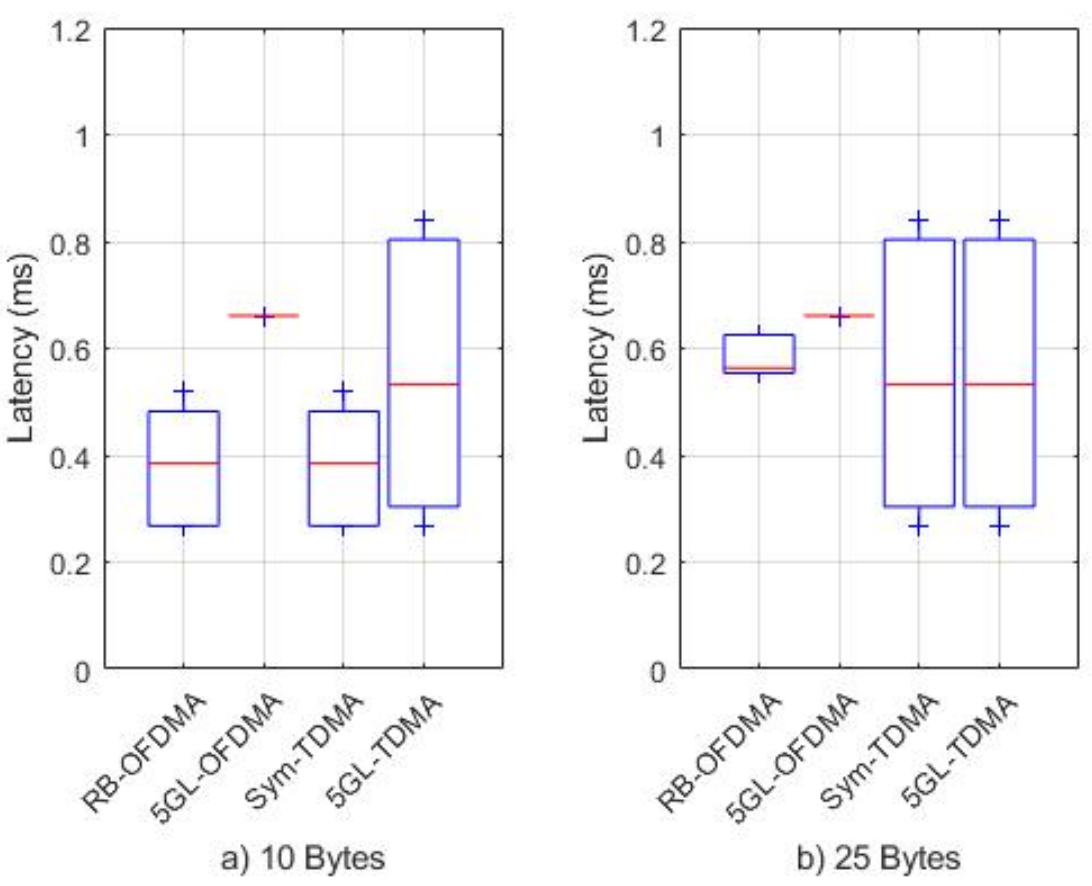

**Figure 10**. UL latency experienced as a function of the packet size (BW=20MHz).

## 7.3. Impact of the MCS

Now, we evaluate the impact of the MCS used for the packet transmissions. Figure 11 shows the maximum UL latency experienced by the UEs and the percentage of allocated radio resources using CG with 5GL-TDMA, 5GL-OFDMA, Sym-OFDMA and RB-OFDMA when MCS 12, 20 and 28 are used and *BW*=20 MHz and packets of 10 bytes are transmitted. The percentage of radio resources allocated is calculated with respect to the number of radio resources used by 5GL-TDMA. Figure 11.a shows that Sym-OFDMA and RB-OFDMA reduce the maximum latency experienced by the UEs for all the evaluated MCSs compared with 5GL-OFDMA and 5GL-TDMA. This is thanks to the most efficient use of radio resources, as shown in Figure 11.b. When MCS increases, a lower number of RBs are demanded by each UE to transmit their packets. Sym-OFDMA and RB-OFDMA adjust the number of allocated resources to the UEs demand exploiting the flexibility offered by OFDMA. However, 5GL-TDMA and 5GL-OFDMA maintain the same radio resource allocation for all MCS values evaluated due to the constraints introduced by the multiple access schemes implemented in 5G-LENA. As a result, Sym-OFDMA and RB-OFDMA reduce the number of allocated radio resources and the maximum experienced UL latency compared with 5GL-TDMA and 5GL-OFDMA. For example, Sym-OFDMA and RB-OFDMA reduce the maximum UL latency by 48.63% and 43.22%, respectively, compared with 5GL-OFDMA when MCS 28 is used. These results are achieved using only 80.7% and 66.7% of the radio resources used by 5GL-OFDMA. Compared with 5GL-TDMA, Sym-OFDMA and RB-OFDMA reduce the maximum UL latency 59.5% and 55.3%, respectively, using only 15.38% and 26.67% of the radio resources.

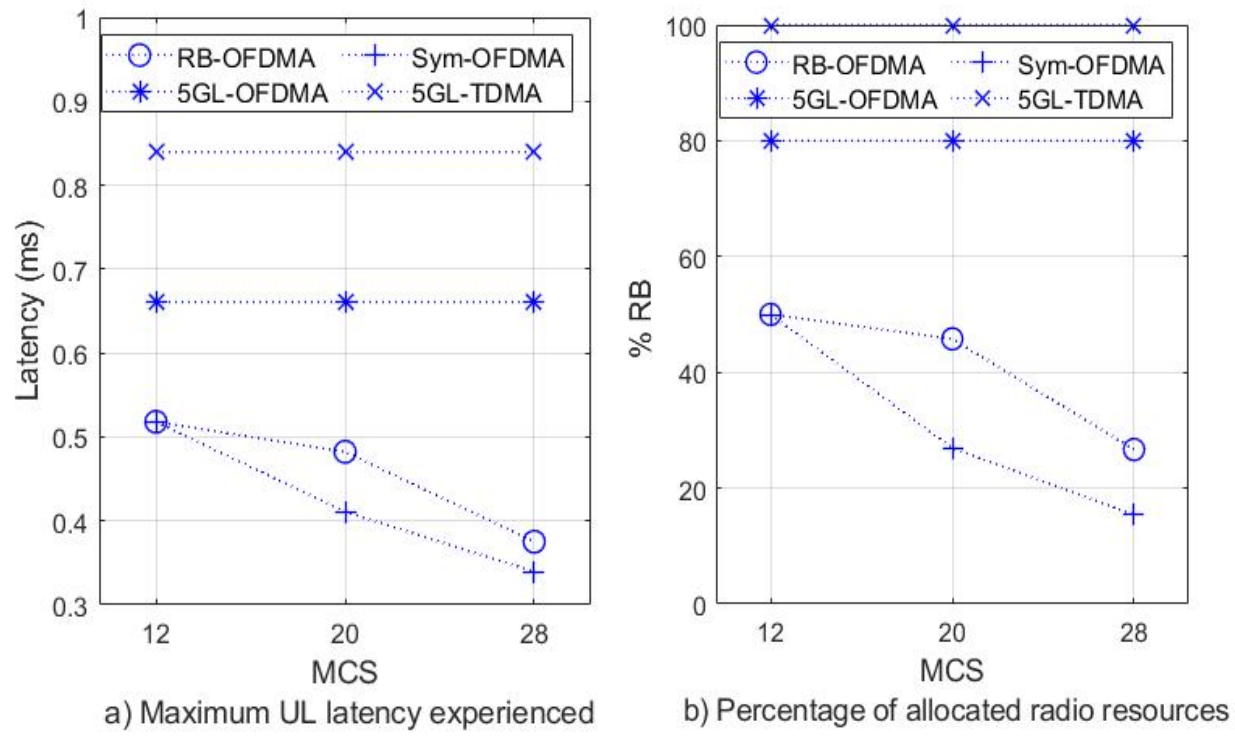

**Figure 11.** Performance as a function of the MCS ($BW$ = 20 MHz, packet size = 10 bytes).

### 7.4. Impact of the frame and slot format

Finally, we evaluate the performance achieved using CG with 5GL-TDMA, 5GL-OFDMA, Sym-OFDMA and RB-OFDMA when only part of the OFDM symbols within a single slot is reserved for UL traffic. To this end, we evaluate the use of 3 different TDD frame and slot configurations. In particular, we consider that the first 1, 5, and 9 OFDM symbols within a slot of a frame are used for DL transmissions, and the last 13, 9, and 5 OFDM symbols of each slot are used for UL transmissions; the three configurations are referred to as 1D13U, 5D9U, and 9D5U, respectively[5]. Figure 12 shows the boxplot of the UL latency experienced by UEs when different frame and slot configurations are used, $BW$=20 MHz, and packet size is equal to 10 bytes. Figure 12 shows that, as expected, the latency increases for all multiple access schemes and scheduling policies when the number of OFDM symbols reserved for UL transmissions decreases. Sym-OFDMA and RB-OFDMA always provide the lowest latency values thanks to the higher flexibility offered by OFDMA multiple access scheme. 5GL-TDMA is the one for which the experienced UL latency increases more. This is because 5GL-TDMA requires a larger number of slots to serve all the UEs as the number of OFDM symbols reserved for UL within a slot decreases.

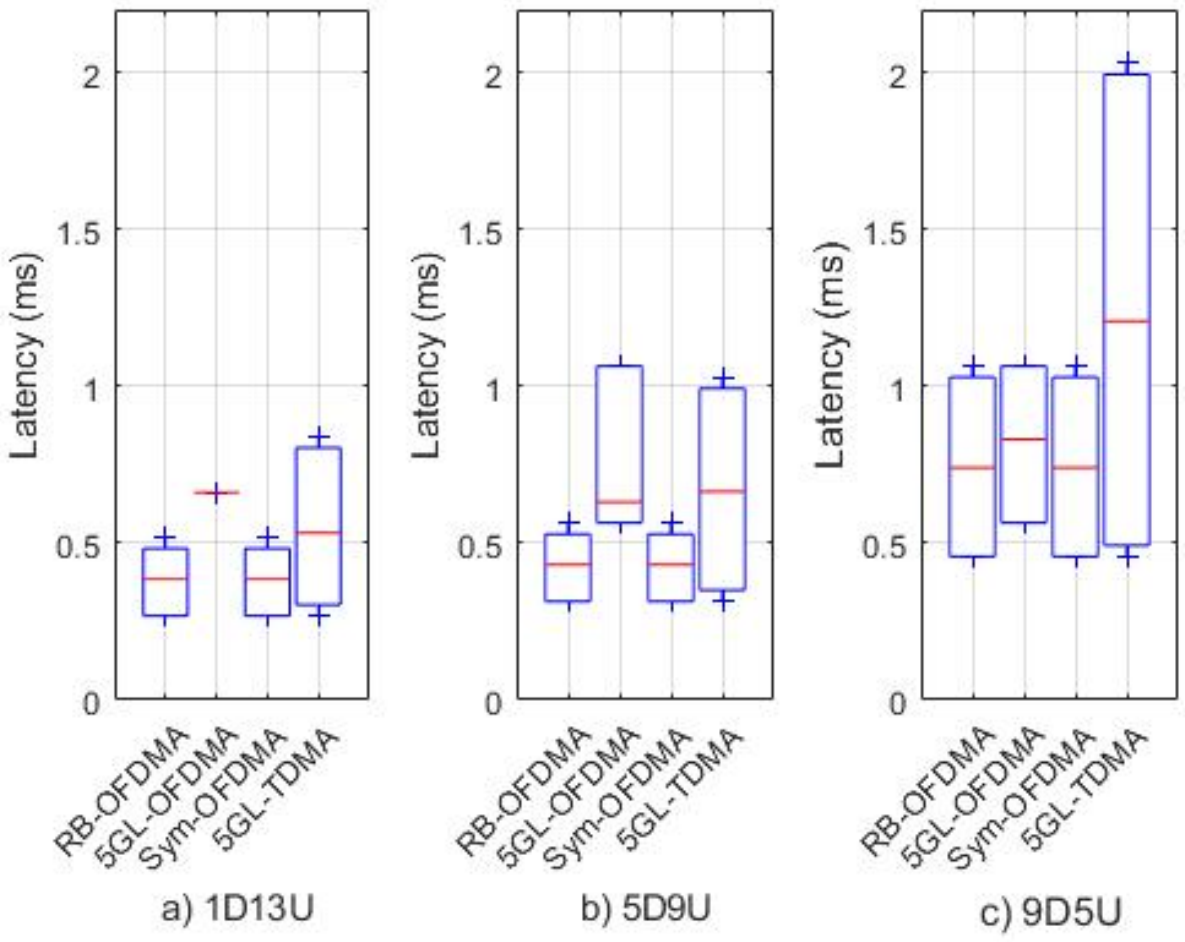

**Figure 12.** Latency experienced with the different schemes as a function of the slot format ($BW$ = 20 MHz, packet size = 10 bytes).

## 8. Conclusions

This paper has presented a detailed description of the first implementation of configured grant scheduling in an open-source 5G NR simulator (to the best of the authors' knowledge), in particular in 5G-LENA. The code of configured grant is publicly available in [7]. Configured grant pre-allocates radio resources to the UEs and avoids the signaling exchange between the UE and the gNB to request/inform about the allocated radio resources reducing the latency of the transmission. Configured grant is key for the support of time-critical services in 5G networks. This work is a valuable contribution since the availability of simulation tools that accurately model all the functionalities of 5G NR and, in particular, configured grant, is fundamental for the research in 5G and beyond networks supporting time-critical services. In particular, we have presented the new functionalities and modifications included at the MAC and PHY layers of 5G-LENA to integrate configured grant in the simulator. To accurately model the high flexibility offered by 5G NR in the radio resource allocation process, we have also implemented the 5G NR OFDMA in the simulator that allows radio resources to be shared simultaneously in time and frequency by different UEs. Using OFDMA multiple access scheme, it is possible to transmit using any number of

[5] We consider that several devices generate DL traffic in the system. The results showed that the performance experienced by the UL traffic is not affected by the amount of traffic generated in DL, but by the number of radio resources reserved for the transmission of UL traffic and the frame and slot configuration.

OFDM symbols. The latency results achieved with configured grant in 5G-LENA match with the latency values reported in previous analytical studies, which validates the implementation of configured grant in 5G-LENA.

We have also implemented two scheduling policies that are applied with configured grant and OFDMA to demonstrate the flexibility and capabilities of 5G NR to support time-critical services. The proposed scheduling policies exploit the flexibility offered by OFDMA to guarantee low latencies and efficient use of radio resources. We have considered a case study where a 5G NR cell covers an industrial scenario where a closed-loop control application demands low latency communications. The results have shown that the use of CG with the proposed scheduling policies and OFDMA reduces the maximum latency experienced by UEs by up to 48.63% compared to the latency experienced when the multiple access schemes previously implemented in 5G-LENA are used. This result is achieved using 80.7% fewer radio resources. The results have shown that the use of scheduling policies that make efficient use of radio resources is more critical the smaller the size of the packets to be transmitted.

**Acknowledgments**

This work has been funded by MCIN/AEI/10.13039/501100011033 through the project PID2020-115576RB-I00, and by European Union's Horizon Europe Research and Innovation programme under Grant Agreement No 101057083.